\documentclass[journal]{IEEEtran}
\ifCLASSINFOpdf
\else
\fi
\usepackage{graphicx}
\usepackage[x11names,table,xcdraw]{xcolor}
\usepackage{epstopdf}
\usepackage[caption=false]{subfig}
\usepackage{multirow}

\begin{document}
%
\title{A Multicriteria Evaluation for Data-Driven Programming Feedback Systems: Accuracy, Effectiveness, Fallibility, and Students' Response}
%
%
%

\author{Preya~Shabrina,
        Samiha~Marwan,
        Andrew~Bennison,
        Min Chi,
        Thomas Price,
        and~Tiffany~Barnes,~\IEEEmembership{~Fellow,~IEEE}
\IEEEcompsocitemizethanks{\IEEEcompsocthanksitem Preya Shabrina, PhD Student, Department
of Computer Science, North Carolina State University, Raleigh,
NC, 27695.\protect\\
E-mail: pshabri@ncsu.edu}
\thanks{Manuscript received April 19, 2005; revised August 26, 2015.}}

%
%

\markboth{IEEE Transactions on Learning Technologies,~Vol.~XX, No.~X, XXXXXX~20XX}%
{Shell \MakeLowercase{\textit{et al.}}: Bare Demo of IEEEtran.cls for IEEE Journals}
%



\maketitle

\begin{abstract}
Data-driven programming feedback systems can help novices to program in the absence of a human tutor. Prior evaluations showed that these systems improve learning in terms of test scores, or task completion efficiency. However, crucial aspects which can impact learning or reveal insights important for future improvement of such systems are ignored in these evaluations. These aspects include inherent fallibility of current state-of-the-art, students' programming behavior in response to correct/incorrect feedback, and effective/ineffective system components. Consequently, a great deal of knowledge is yet to be discovered about such systems. In this paper, we apply a multi-criteria evaluation with 5 criteria on a data-driven feedback system integrated within a block-based novice programming environment. Each criterion in the evaluation reveals a unique pivotal aspect of the system: 1) How accurate the feedback system is; 2) How it guides students throughout programming tasks; 3) How it helps students in task completion; 4) What happens when it goes wrong; and 5) How students respond generally to the system. Our evaluation results showed that the system was helpful to students due to its effective design and feedback representation despite being fallible. However, novices can be negatively impacted by this fallibility due to high reliance and lack of self-evaluation. The negative impacts include increased working time, implementation, or submission of incorrect/partially correct solutions. The evaluation results reinforced the necessity of multi-criteria system evaluations while revealing important insights helpful to ensuring proper usage of data-driven feedback systems, designing fallibility mitigation steps, and driving research for future improvement. 
\end{abstract}


%
\IEEEpeerreviewmaketitle

\section{Introduction}
\IEEEPARstart{B}lock-based programming environments are intended to provide novices with the ability to engage in motivating, open-ended, and creative programming tasks with features that limit syntax errors but allow for simplified programming for interactive media~\cite{garcia2015beauty, dann2012mediated}. These environments are often equipped to provide automated support such as misconception-driven feedback~\cite{Gusukuma2018}, next-step hints ~\cite{marwan2019evaluation}, or adaptive feedback on task or sub-task completion~\cite{Marwan2020ICER}, which have been shown to improve students' engagement and learning. Recently, data-driven automated programming hints and feedback are being explored by researchers as they can be generated automatically using historical or current log data with reduced engagement of experts~\cite{zhitoward, price2017hint, Marwan2020ICER, price2016generating}. Our prior study showed that data-driven adaptive feedback increased students' engagement with programming tasks, and improved their programming performance~\cite{Marwan2020ICER}. Existing literature also showcases studies that demonstrate the positive impact of such feedback on students’ performance or learning [For example,~\cite{marwan2019evaluation},~\cite{Gusukuma2018}, etc.]. 

Note that evaluating the correctness of a program automatically to provide feedback is a complex task and the state-of-the-art is yet to be perfected. Data-driven feedback systems usually provide feedback by comparing previous students' correct solutions with the program under evaluation. These systems often do not have a way to adapt to new programming approaches, making such feedback systems inherently fallible. Michael Ball~\cite{ball2018lambda} designed a autograder based feedback system for Snap (a block-based programming environment designed for novices) that gives feedback based on detected correct code blocks (draw square, repeat, etc.). Although the autograder was able to reduce staff workload and guided students to finish assignments in the absence of TAs, the author mentioned it not to be precise enough yet. However, prior evaluations of data-driven programming feedback systems found in literature have failed to consider this crucial fact; this has resulted in inaccurate reliability measurements and knowledge of data-driven feedback systems, and how students’ programming behaviors are impacted when the system \textit{fails} to provide correct feedback. Also, missing are insights into how these systems are \textit{succeeding} in helping students despite providing inaccurate feedback. Therefore, we concluded that existing evaluation methods for data-driven programming feedback systems are insufficient and missing important insights.  

In this paper, we propose an evaluation mechanism for data-driven programming feedback systems that focuses on analyzing students' programming behavior in relationship with the effectiveness, correctness, and fallibility of system features. The evaluation consists of 5 criteria: 1) How accurate the feedback system is; 2) How it guides students throughout programming tasks; 3) How it helps students in task completion; 4) What happens when it goes wrong; and 5) How students respond generally to the system. The criteria are formalized in Section \ref{sys_evaluation}. We applied this evaluation method to a data-driven adaptive immediate feedback (\textbf{DDAIF}), integrated within a block-based programming environment, \textbf{iSnap}. Our methodologies for the evaluation involve expert analyses of student code traces (to analyze and quantify system performance/fallibility), graph-based visualizations of student problem solution paths (to demonstrate how the system drives students' programming problem-solving attempts), and case studies (to demonstrate specific instances where the system was helpful or misleading). The results of our evaluation not only give important insights on our system signifying the necessity of a multi-criteria evaluation mechanism like ours but also gives direction towards the improvement of such systems and motivate research for an effective approach to communicate the fallibility of data-driven feedback systems to ensure a smooth learning process for novice programmers.

\section{Related Work}
Several block-based programming environments have been designed to reduce the difficulties students face while learning a new programming language in various ways. For example, Alice~\cite{dann2012mediated}, and Snap\textit{!}~\cite{garcia2015beauty} provide drag-and-drop coding and immediate visual code execution. Scratch focuses on allowing novice programmers to create tinkerable projects of their interest~\cite{resnick2009scratch} while promoting peer collaboration. App Inventor for Android, developed by MIT, is a powerful block-based programming environment that facilitates the development of mobile applications with real-world utilities reducing the initial barrier of learning programming~\cite{morelli2011can}. Each of these environments shares a goal of simplifying a programming environment without the potential for syntax errors ~\cite{dann2012mediated}. Research has shown that these programming environments are more engaging in terms of reduced idle time while solving a programming problem~\cite{price2015comparing} and can produce positive learning outcomes in terms of grades ~\cite{dann2012mediated} and the number of goals completed in a fixed amount of time~\cite{price2015comparing}.

To provide novice students with individualized tutoring support, researchers have integrated data-driven intelligent features into block-based programming environments. These intelligent features dynamically adapt teaching support to mitigate personalized needs~\cite{murray1999authoring}. For example, iSnap~\cite{price2017isnap} is an extension of Snap\textit{!} that provides on-demand hints generated from students' code logs using the Source Check Algorithm~\cite{Price2017EDM}. Gusukuma et al.~\cite{Gusukuma2018} integrated automatic feedback based on learners’ mistakes and underlying misconceptions into BlockPy~\cite{bart2017blockpy}, and showed that it significantly improved students’ performance. Rivers and Koedinger’s ITAP, a data-driven tutor for an introductory Python course, was able to create hints that would lead to a solution from 98\% of incorrect solutions ~\cite{rivers2017data}. Such data-driven approaches are being integrated to provide more automated adaptive tutoring support in novice programming environments. For example, iSnap showcased the first attempt to integrate data-driven support into a block-based programming environment. Zhi et al.~\cite{zhitoward} proposed a method of generating example-based feedback from historical data for iSnap. They extracted correct solution features from previous students' code and used those features to remove extraneous codes from current student code and produced pairs of example solutions that were provided on an on-demand basis.

The impact and effectiveness of various tutoring supports and data-driven intelligent features integrated into novice programming environments have been explored by researchers from various perspectives. Zhi et al.~\cite{zhi2019exploring} demonstrated the adoption of worked examples in a novice programming environment and found out that worked examples helped students to complete more tasks within a fixed period of time, but not significantly more. Price et al.~\cite{price2017hint} explored the impact of the quality of contextual hints generated from students' current code on students' help-seeking behavior. They found out that students who usually used hints at least once performed as good as students who usually do not perform poorly and also the quality of the first few hints is positively associated with future hint use and correlates to hint abuse. Marwan et al.~\cite{marwan2019evaluation} evaluated the impact of data-driven automated programming hints on students' performance and learning and argued that automated hints improved learning on subsequent isomorphic tasks when accompanied by self-explanation prompts. Corbett and Anderson demonstrated with their Lisp tutor that students learn more efficiently when the tutor algorithm has better control over the provided feedback ~\cite{corbett2001locus}. Mao et al. explored machine learning based models and approaches (Bayesian Knowledge Tracing/Long Short Time Memory~\cite{mao2018deep}, recent temporal patterns~,\cite{mao2019one} etc.) to identify the need for interventions. To evaluate the approaches, they depended on traditional accuracy metrics (accuracy, F1-score, AOC/ROC, etc.). However, they did not further evaluate what happens when a decision for intervention is taken for a user based on an incorrect output from their model. Dong et al.~\cite{dong2021using} proposed a data-driven \emph{SourceCheck}~\cite{price2017evaluation} method to detect students' struggles and need for intervention. They used an expert-opinion based accuracy for evaluation without a user-impact analysis.  

Although the effectiveness of data-driven programming support or feedback systems is heavily researched, the limitations and fallibility of such systems and students' responses to system features remain unknown and under-explored. Also, the effectiveness evaluations are mostly based on learning measured by surface-level metrics like test scores or task completion rates. A deeper evaluation of such systems focusing on user impact is yet to be done which could reveal how students' learning and programming experience are impacted by the system. Thus, in this paper, we applied a multi-criteria evaluation on DDAIF system to shed light on the yet unrevealed important insights on the family of data-driven feedback systems that could be potentially helpful to get future direction on methods to improve and better exploit the potential of these systems.
\section{System Design}
\subsection{The Novice Programming Environment}\label{iSnap}
We built the data-driven adaptive immediate feedback system (\emph{DDAIF}) in iSnap~\cite{price2017isnap}, a block-based intelligent novice programming environment. Within iSnap, students can drag and drop code statements called blocks to formulate a problem solution. Students can observe visual outputs as they execute their program. This environment also provides students with on-demand programming hints, and can sequentially log students' code edits (e.g. adding or deleting a block) while programming as \textit{code traces}. This logging feature allows researchers or instructors to replay all students' edits in the programming environment, and detect the time for each edit.

\subsection{Generating Data-Driven Adaptive Immediate Feedback (\emph{DDAIF})}\label{DDAIF}
\begin{figure}
\centering
\includegraphics[width=0.5\textwidth]{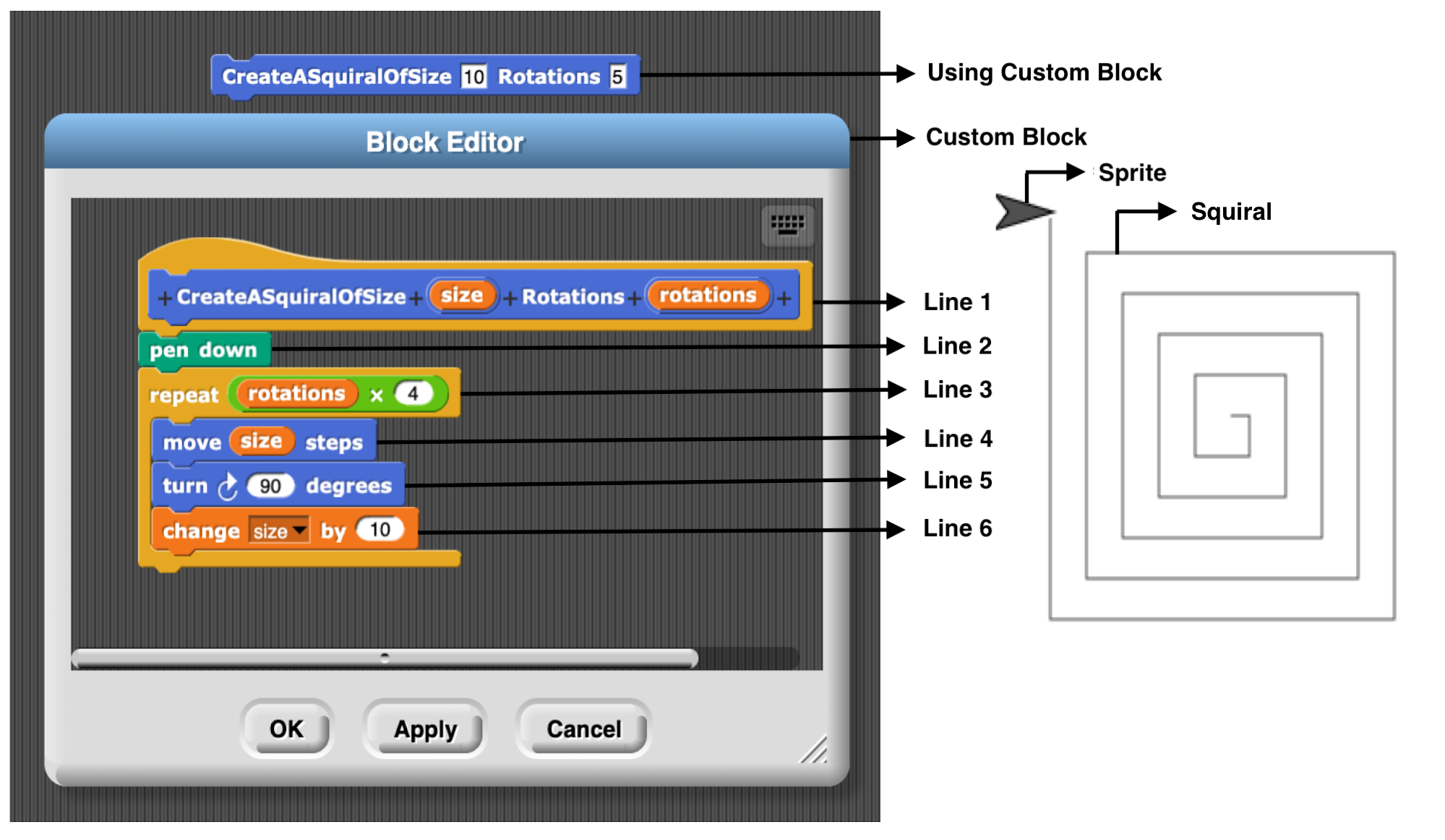}
\caption{a) A sample expert solution to solve the Squiral assignment; and b) Expected output on the Right.}\label{fig:expert_solution_output}
\end{figure}

\begin{figure}
\centering
\includegraphics[width=0.5\textwidth]{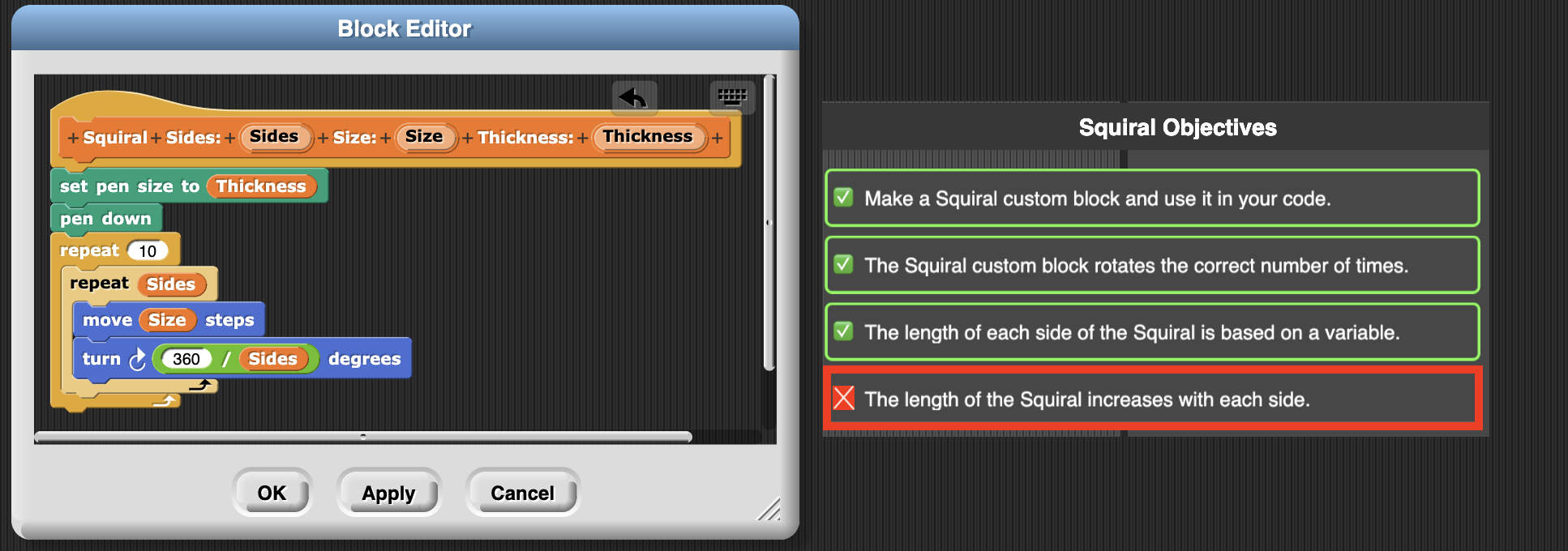}
\caption{DDAIF Interface with 4 Objectives for Squiral problem and Adaptive Immediate Feedback based on Correct and Incorrect Objective Detections.}\label{fig:demonstration}
\end{figure}

We call our generated feedback \textit{data-driven}: generated from historical correct student solutions to programming problems, \textit{adaptive}: given feedback is adapted to individual student solutions, and \textit{immediate}: feedback is given immediately based on each code edit. We implemented and presented DDAIF to students while they solved a programming problem called \textit{Squiral} as a homework assignment. We first describe the \textit{Squiral} problem and then the problem's feedback generation process.\\
\textbf{The Squiral Problem:} \emph{Squiral} is a programming problem that asks to construct a program to draw a spiral square-like shape. One solution and its corresponding output is shown in Figure \ref{fig:expert_solution_output} and described below:\\
\textit{Custom block:} A \textit{custom block} (similar to a function/method in textual programming languages), named `CreateASquiralOfSize', is created and used that contains the logic to draw a \textit{Squiral} and moves the \textit{Sprite} in the expected way.\\ 
\textit{Line 1:} The \textit{custom block} takes two parameters: `size' indicating the length of a side of the innermost square, and `rotation' indicating the number of square loops to draw.\\
\textit{Line 2:} The `pen down' block [Line 2] effectively performs the drawing.\\
\textit{Line 3:} A loop that repeats \textit{Line 4-6} to draw one square loop in each iteration.\\
\textit{Line 4:} Draws one side of a square of length `size'.\\
\textit{Line 5:} Turns the sprite by 90 degrees to the right\\
\textit{Line 6:} Changes the value of `size' by 10 to prepare for drawing the next side of the square.
\\\textbf{Data Collection for Feedback Generation:} The iSnap platform has been used by novice/non-cs majors students taking a CS0 course at a public university in the US for in-class and homework programming assignments. The code traces of different student solution attempts for specific problems were collected in the Spring and Fall semesters for the years 2016-17. From this code data pool, correct student solutions to the \textit{Squiral} problem were filtered out using their final assignment grades. We used these data to generate adaptive immediate feedback for the corresponding problems.\\
\textbf{Algorithm to Identify Common Correct Solution Features:} DDAIF is given based on common features (sequence of code blocks with a particular functionality) of previous students' correct solutions to a problem. For example, a custom block, or code blocks to draw a square could be features of correct solutions for the \textit{Squiral} problem. To identify common features of correct students solutions of \textit{Squiral}, we used the data-driven algorithm described in ~\cite{zhi2018reducing} which is summarized below:\\
\textit{Represent solutions:} Represent each correct student solution as an abstract syntax tree(AST).\\
\textit{Construct initial feature set, C, containing all code shapes:} From the ASTs, extract code shapes represented as pq-Grams~\cite{augsten2005approximate} by starting from each node, and by including its (p-1) immediate ancestors and a maximum of q of its children for all $p \in 1...3$ and $q \in 1...4$.\\
\textit{Remove redundant code shapes:} For each code shape $c \in C$, construct $S_{c}$, a set of code snapshots containing $c$. Then, for each pair of code shapes $\{ci, cj\}$, calculate Jaccard similarity (measure of similarity due to co-occurrance), $J_{ij}$ as ($\|S_{ci} \cap S{cj} \|/\|S_{ci} \cup S_{cj} \|$). If $J_{ij} > 0.975^2$, remove from C the smaller code shape in the pair.\\
\textit{Construct Decision Shapes:} For each $c \in C$, calculate support, $Su_c$, fraction of correct solutions containing $c$. For each pair of code shapes $\{ci, cj\}$, calculate overlap, $O_{ij}$, fraction of correct solutions containing both $C_i$, and $C_j$. If for $c_i$, $O_{ij}$ is the smallest and $Su_{cj} < Su_{ci}$,  construct decision, $d = c_i \cup c_j$ and add it to C. Remove from C, any code or decision shapes with support less than $0.9^2$. Decision shapes represent alternate features in different strategies.\\
\textit{Hierarchical clustering to generate a final set of features:} At each iteration of clustering, combine two most Jaccard similar (most co-present) code/decision shapes in C as a single feature. Stop when there is a steep drop in resolution in ``Elbow Plot''. Resolution is defined as the ability of a feature set to differentiate between states of subsequent snapshot and its value decreases as shapes are combined.\\
\textbf{Formulating Objectives from Features:} After extracting the features for the Squiral problem from prior student solutions, two experts in block-based programming grouped the features to formulate subgoals or objectives described in natural language for new students attempting to solve Squiral. The objectives along with their natural language labels and required features are presented in the left and right columns of Table \ref{tab:obj_require} respectively. For example, Objective 1 states `Make a Squiral custom block and use it in your code' which requires feature 1 ($F_{1}$ : presence of a custom block) for completion. In Figure \ref{fig:demonstration}, line 2 corresponds to Objective 2, line 4 completes Objective 3, and lines 2 and lines 4-6 correspond to Objective 4. By applying the algorithm described previously, we extracted 7 features for the Squiral problem that were grouped to form 4 key, concrete  objectives. The natural language labels of the objectives were reviewed by experts of the programming course verifying their interpretability and representability of the Squiral problem. These objectives were presented to new students to accomplish while they solved the same problem.
\begin{table}[h!]
\centering
\caption{Sample requirements to complete each objective.}
\label{tab:obj_require}
\begin{tabular}{ll}
\hline
Objective Number and Label                                                                                                                                                         & Required Features for  Completion                                                                                                                                          \\ \hline
\begin{tabular}[c]{@{}l@{}}1: Make a Squiral custom block\\     and use it in your code.\end{tabular} & $F_{1}$: Create and use a custom block                                                                                                                                                \\ \hline
\begin{tabular}[c]{@{}l@{}}2: The Squiral custom block \\     rotates the correct number of\\     times.\end{tabular}                                                              & \begin{tabular}[c]{@{}l@{}}$F_{2}$: A loop as follows:\\ \textbf{\textit{repeat y * z}}\\ Or a nested loop\\ \textbf{\textit{repeat y}}\\    \textbf{\textit{....repeat z}}\\ {[}y = rotation count; z = 4{]}\end{tabular}                           \\ \hline
\begin{tabular}[c]{@{}l@{}}3: The length of each side of the\\     Squiral is based on a variable.\end{tabular}                                                                    & \begin{tabular}[c]{@{}l@{}}$F_{3}$: Within loop: \textbf{\textit{move x steps}}\\ {[}variable x = length of a side{]}\end{tabular}                                                                             \\ \hline
\begin{tabular}[c]{@{}l@{}}4: The length of the Squiral \\     increases with each side.\end{tabular}                                                                              & \begin{tabular}[c]{@{}l@{}}$F_{4}$: Outside loop: \textbf{\textit{pen down}}\\ And Within loop:\\        $F_{5}$: \textbf{\textit{move x steps}}\\        $F_{6}$: \textbf{\textit{turn 90 degrees}}\\        $F_{7}$: \textbf{\textit{change x by some\_value}}\end{tabular} \\ \hline
\end{tabular}
\end{table}
\\\textbf{DDAIF Interface:} We designed the DDAIF interface based on our prior work~\cite{Marwan2020ICER}. This interface (Fig. \ref{fig:demonstration}) includes a progress panel that displays the 4 objectives needed to complete Squiral constructed from the data-driven features extracted using the method described in Section \ref{DDAIF}. Our system further provides adaptive immediate feedback based on the completion of these objectives.\\  
\textit{Feature Detection and Objective Completion Based Adaptive Immediate Feedback:} After a student makes an edit while programming, the DDAIF system converts the current code snapshot into an abstract syntax tree (AST). From this AST, the system generates a sequence of zeroes(0) and ones(1) called a feature state (e.g. 1100000, where the first two ones indicate the presence of the first two features [$F_{1}$ and $F_{2}$], and zeroes indicate an absence of rest of the features). Initially, all the objectives in the progress panel are deactivated. Once the system detects the presence of all features required to complete an objective, the color of the progress bar representing that objective changes to green. If the system detects the absence of a feature that was present before (i.e. a broken feature), its corresponding objective turns red as depicted in Figure \ref{fig:demonstration}. This adaptive feedback is given immediately after a code edit is made, and is solely based on a student's current code state.
\section{Procedure}
In the Spring of 2020, We deployed iSnap to be used for in-class programming assignments in an introductory computing course in a public research university in the United States. However, the students used iSnap (\ref{iSnap}) coupled with \emph{DDAIF} while solving the \textit{Squiral} problem as a homework assignment. 27 non-CS majors participated in the course and in this study. The data we collected consists of code snapshots for every edit in student code while solving the \textit{Squiral} problem with corresponding timestamps and received feedback. Afterward, we defined an evaluation framework consisting of 5 criteria and developed data-driven methodologies (that use only our collected data) for each of those criteria as we evaluated the DDAIF system using the framework. Our methodologies involve quantitative, exploratory, and case-study based analyses. 
1) How accurate the feedback system is; 2) How it guides students throughout programming tasks; 3) How it helps students in task completion; 4) What happens when it goes wrong; and 5) How students respond generally to the system.
\subsection{The Multicriteria Evaluation Mechanism for Data-Driven Feedback Systems}\label{sys_evaluation}
We formalize here the 5 criteria of our evaluation framework for data-driven feedback systems:
\begin{itemize}
\item \textbf{\emph{Evaluation Criterion 1(Computational Accuracy):}} How accurate the feedback system is.
\item \textbf{\emph{Evaluation Criteria 2(Guidance during Programming Tasks):}} How it guides students throughout programming tasks.
\item \textbf{\emph{Evaluation Criteria 3(Effective Feedback Components):}} How it helps students in task completion. 
\item \textbf{\emph{Evaluation Criteria 4(Impact of Fallibility):}} What happens when it goes wrong.
\item \textbf{\emph{Evaluation Criteria 5(Generalized Student Response):}} How students respond generally to the system.
\end{itemize}
The first criterion of this evaluation framework, computational accuracy, provides initial insights into the potential or limitations of the system. This criterion also guides later evaluation steps in the right direction (for example, accuracy less than $100\%$ indicates a necessity of fallibility analysis). However, our observations suggest that for data-driven feedback systems only accuracy is not enough to measure usability. Therefore, we introduced the later four criteria to answer how a fallible data-driven feedback system guides/helps students in programming tasks and how the fallibility impact students' behavior. In the subsequent sections, we describe data-driven methods that we used to evaluate our DDAIF system based on the 5 criteria and report the insights from the evaluation results.


\section{Evaluation Criterion 1: Computational Accuracy}\label{evaluation_c1}
\subsubsection{Method}
To measure the computational accuracy of the DDAIF system, we conducted a traditional system evaluation where we calculated how often the system was correct or incorrect in giving feedback. From students' sequential code trace of solving Squiral, we (the researchers) filtered out all instances when DDAIF provided feedback based on the detection of completed/broken objectives, and instances when feedback should have been given but the system did not give any. We tagged each of these cases with one of the following:
\begin{itemize}
\item \textit{True Positive (TP)} : An objective was detected as completed [marked green] by the system and according to researchers, the objective was actually completed at the time of detection. 
\item \textit{True Negative (TN)} : An objective was detected as broken [marked red] by the system and the objective was actually broken at the time of detection according to researchers.
\item \textit{False Positive (FP)} : An objective was detected as completed [marked green] by the system but according to researchers,  the objective was incomplete at the time of detection.
\item \textit{False Negative (FN)} : An objective was completed according to researchers but the system did not detect it or marked it as broken [marked red].
\end{itemize}
 Note that, contrary to the traditional definition of TN, within TN cases we did not include cases when an objective is incomplete (an objective that was never completed before) and undetected by the system. Because, when such cases occur we can not tell for sure if the system actually knew that the objectives were incomplete or if the system is indecisive about the completion of an objective. Also, these cases are frequent, since students are continuously making edits and a lot of times objectives are incomplete. Thus, the inclusion of these cases may introduce uncertainty and overshadow the contribution of certain cases to the accuracy measure.
 After the tagging process, using the count of TP, TN, FP, and FN occurrences, we calculated accuracy, precision, recall, $F_1$ score, true negative rate (TNR), false positive rate (FPR), and false negative rate (FNR) using conventional formulae.
\subsubsection{Computational Accuracy of DDAIF}
While solving Squiral using iSnap with DDAIF, each student received 1 to around 200 detection-based feedback depending on how they approached solving Squiral. With each of those detections tagged with TP, TN, FP, or FN, the DDAIF system achieved a recall/TPR of 83\% (accuracy in detecting completed objectives). However, these correct detections are often (46\% of the time) made slightly early or late. On the other hand, DDAIF can identify incomplete objectives with 61\% accuracy (TNR). This means that the system marks incomplete objectives as complete with high frequency (High false-positive rate of 39\%). However, our system's lower FNR (\textit{17\%}) indicates completed objectives are marked as broken or incomplete less frequently. Overall the system showed an accuracy of 76\%. However, if early or late detections of completed objectives are considered completely inaccurate, DDAIF may show accuracy as low as 55\%.
\subsubsection{Sample Demonstration: Where the System Worked Accurately and Where it Failed}
To reason about the computational accuracy of our system and to identify where our system worked or failed, and why, we further investigated occurrences of correct and incorrect feedback while students solved Squiral. Our observation consistently exhibits that the system failed, when a student showed a new programming approach that was not commonly observed in previous students' code. Here, we describe the code and feedback received by a student as a sample case for demonstration. As shown in figure \ref{fig:demonstration}, our system detected objective 1 accurately when a custom block was created and used in the stage. To detect Objective 2, our system only looked for nested loops (a common feature extracted from previous students’ solutions) and ignored the parameters used in those loops. Thus, in the figure, Objective 2 was incorrectly detected where `10’ and `Sides’ are used as the `repeat’ block parameters. However, according to experts, one of the parameters should be the constant `4’ to indicate the 4 sides of the Squiral. Objective 3 is detected correctly in the code where the `move’ statement used the parameter `Size’ to draw a side of the Squiral. Objective 4 was undetected since a `change’ statement to increase the `Size’ parameter was not used yet.
\subsubsection{Findings}
The computational accuracy of our system and our observation of student codes and received feedback suggest that a data-driven feedback system like ours, developed based on the extraction of common syntactic features from previous students' data and the presence of those features in current students' code, could be occasionally fallible. Since new students often program in new ways, this fallibility may be inherent to such systems. However, our later investigation on how DDAIF guided and helped students in programming tasks despite being fallible (Evaluation Criteria 2 and 3 [Section \ref{sec:cri_2} \& \ref{how_helps}]) aligned with the claim of prior research that these systems are helpful to students. Thus, we recommend computational accuracy of such systems should be used to investigate system limitations or possibly bugs. However, any conclusion on usability should not be drawn on the basis of this evaluation criteria alone.

\section{Evaluation Criteria 2: Guidance During Programming Tasks}\label{sec:cri_2}
This evaluation criterion is introduced to develop a better understanding of how DDAIF guided students throughout Squiral solving attempts and also, to shed light on scenarios when the system failed to guide properly. We adopted a visual state transition diagram based approach to address this criterion. The state transition diagrams represented students' entire problem-solving attempts in terms of completion of objectives. We generated diagrams based on both expert and system detections of complete, incomplete, and broken objectives [Section \ref{sec:STD}]. Note that the expert state transition diagram represents the true picture of students' solution attempts which we used to analyze how the solution attempts were guided or impacted by the system. On the other hand, the system state diagram represents how the system modeled students' solution attempts. Thus, we compared the solution paths extracted from the system state transition diagram against the ones extracted from the expert diagram to understand where the system was different from expert (or human tutor) judgment and could not guide students effectively.

\subsubsection{State Transition Diagrams from Code Trace}\label{sec:STD}
\begin{figure}
\centering
\includegraphics[width=0.2\textwidth]{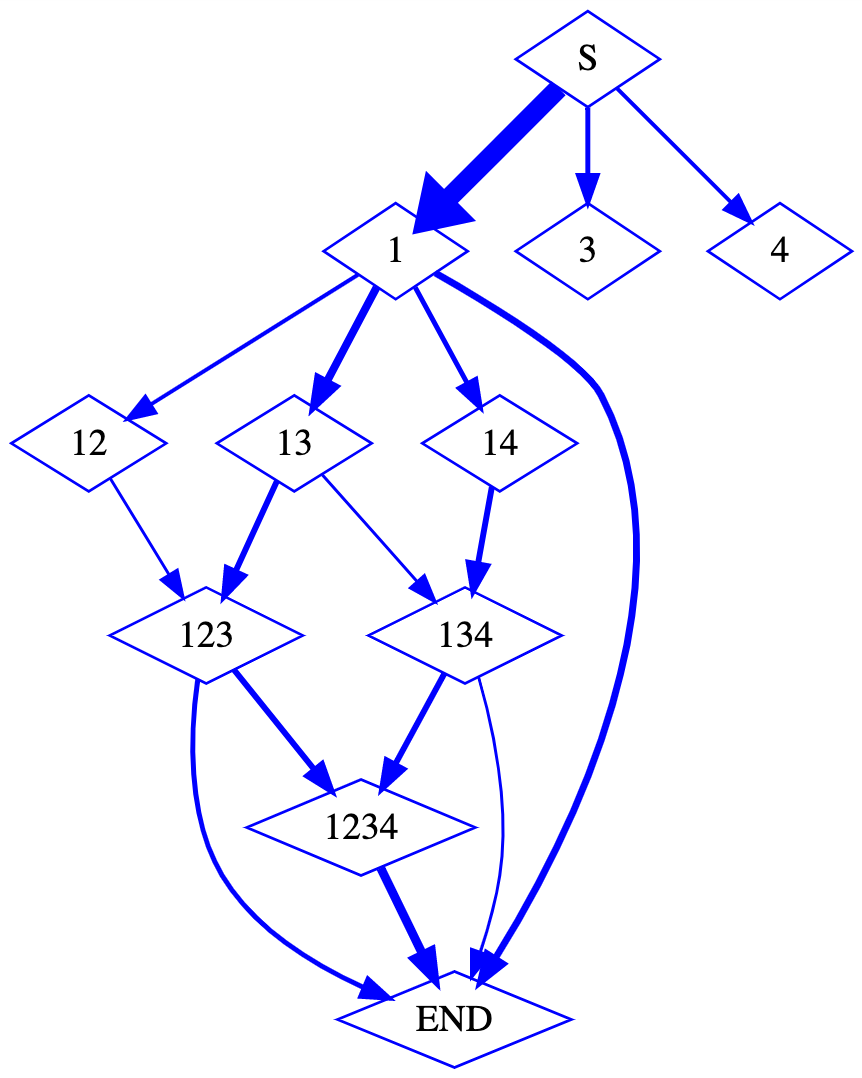}
\caption{State Transition Diagram using System Detections.}\label{fig:sys_detect_diagram}
\end{figure}

\textbf{\textit{Structure:}} The state transition diagrams that we used for our analysis are composed of nodes and edges. The nodes and edges of the diagrams are defined as follows:\\
\textbf{Nodes: }Each node represents a code state defined by the objectives completed at that state. A student's code moves into a new state when one or more objectives are completed or broken. For example, 123$\Rightarrow$13 means the student moved from a state where objectives 1, 2, and 3 are complete to a state where objectives 1 and 3 are complete due to a broken objective 2.\\
\textbf{Edges: }An edge between two nodes or states represents single or multiple code edits that caused the transition. The same state transition can occur due to different code edits.\\
\textbf{\textit{Diagram Notations:}} In the state transition diagrams, nodes are drawn as black oval (expert detected state) or blue diamond shape (system detected state), forward transitions or objective completions are represented with black (expert detected transitions) or blue (system detected transitions) edges, and backward transitions or broken objectives are represented with edges colored red.  Additionally, a node labeled `S’ means `Start’ state, `WC’ means working code (i.e. the code was capable of drawing a Squiral), `NWC’ means a non-working code (i.e. the code was not capable of drawing a Squiral), and `END’ simply indicates the end of an attempt. Note that the system detection based paths end with the `END' node since the DDAIF system cannot tell if a program can draw a Squiral or not. Frequent edges were assigned more weight and drawn with thicker lines.\\\\
\textbf{\textit{Construction of State Transition Diagram:}} To generate a state transition diagram from students' code traces, first, all state transitions found in each student's code traces were documented. Then, occurrences of unique state transitions over 27 students were counted. Using these state transitions, diagrams were generated in the following four phases:\\
 \begin{figure}
\caption{Phases of generating expert detection based state transition diagram.}
\label{fig:phases_diagram}
\begin{center}
\subfloat[Initial State Transition Diagram.]      {
\includegraphics[width=.45\linewidth]{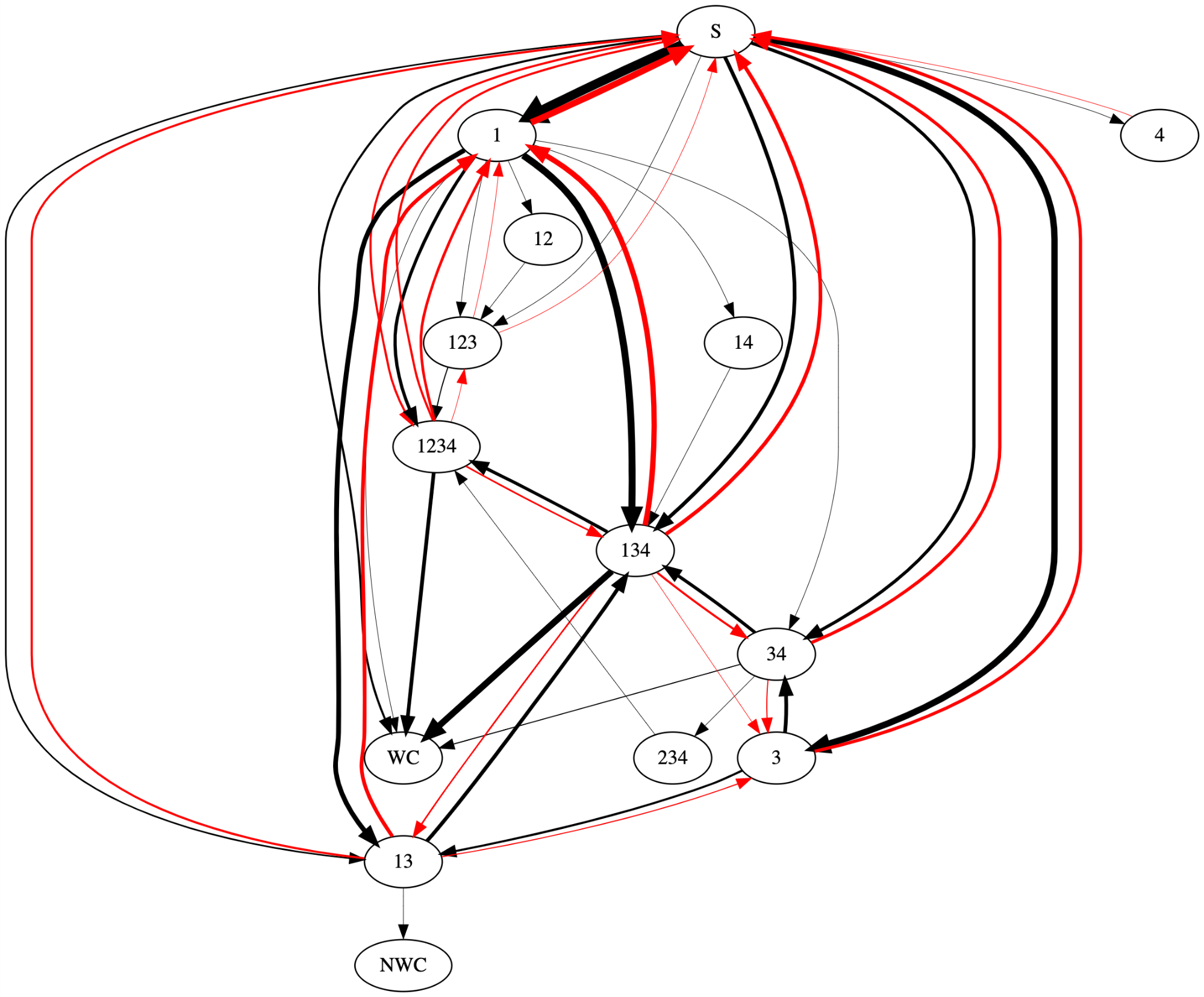}
}\quad
\subfloat[Diagram After Phase 1 Simplification.]      {
\includegraphics[width=.45\linewidth]{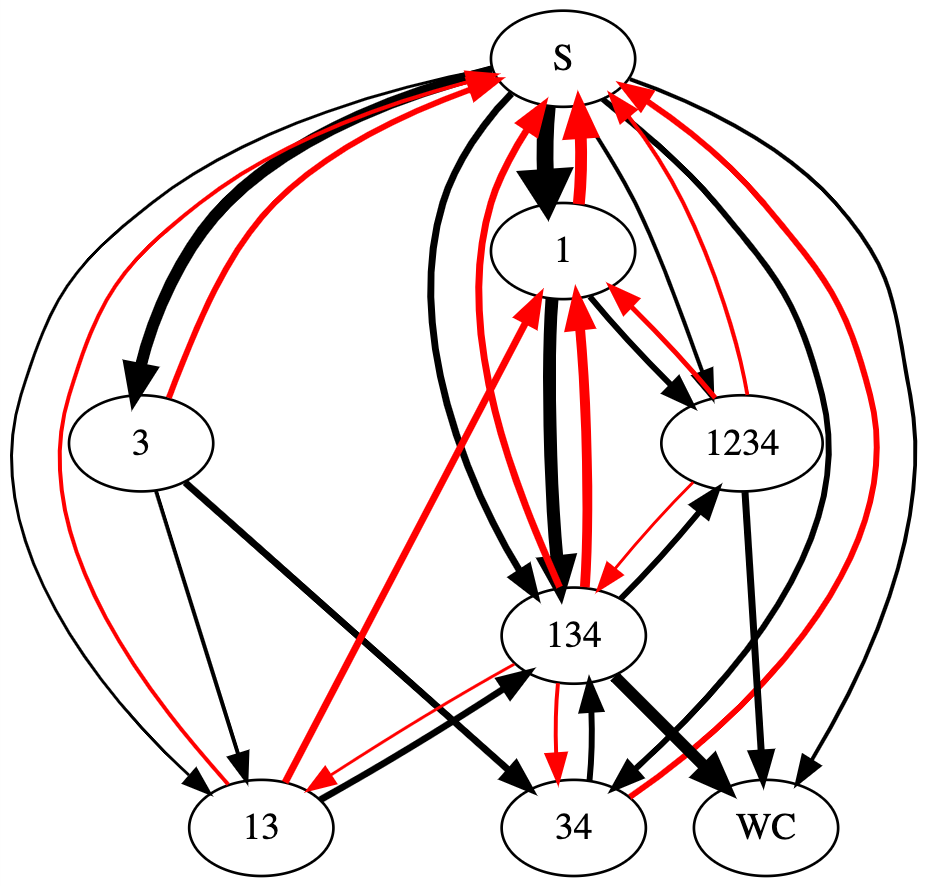}
}\\
\subfloat[Diagram After Phase 2 Simplification.]      {
\includegraphics[height=1.5in, width=.45\linewidth]{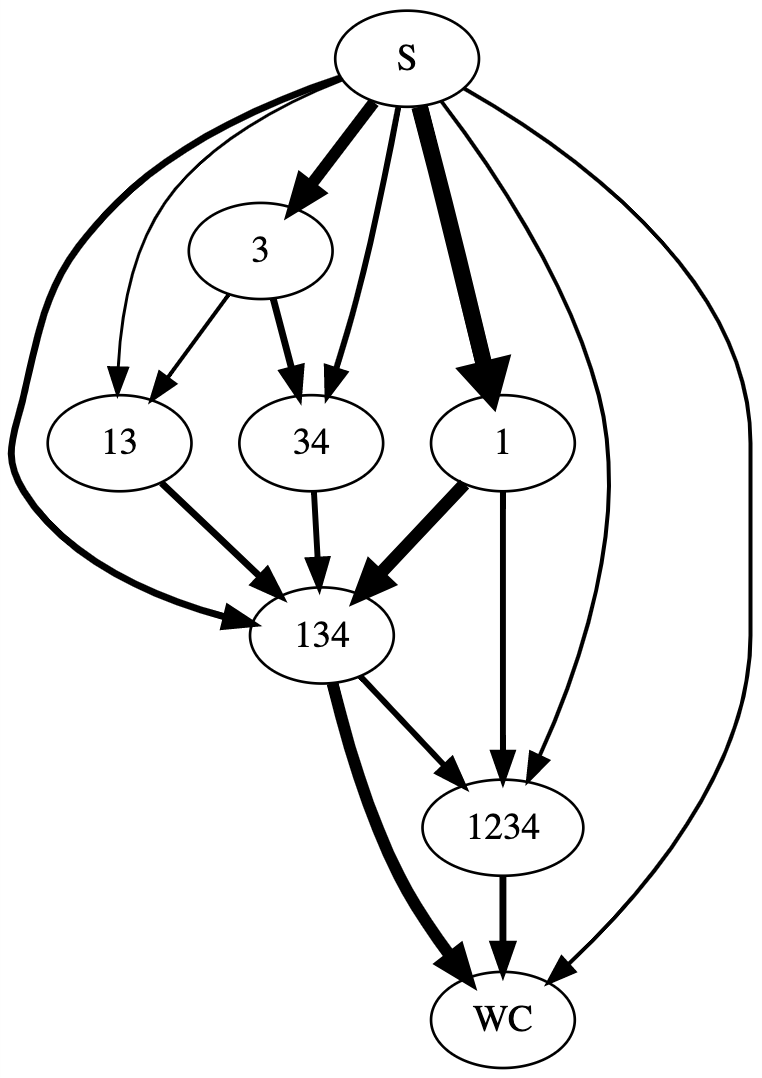}
}\quad
\subfloat[Final Diagram After Phase 3 Simplification.]      {
\includegraphics[height=1.5in,width=.3\linewidth]{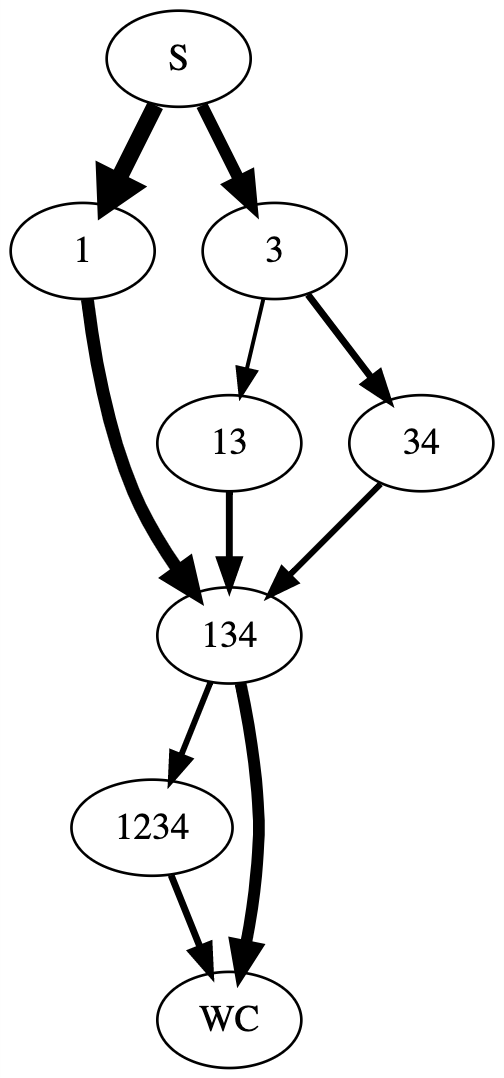}
}
\end{center}
\end{figure}
\textbf{Full diagram:} The initial diagram contains all state transitions that occurred in the solutions of 27 students. It has too many edges and is visually uninterpretable [Figure \ref{fig:phases_diagram}a].\\
\textbf{Simplification Phase 1: }In this phase, we applied a threshold-based reduction to remove infrequent edges. The edges or transitions that occur for only 1-2 students (below 10\% of the total population) were removed [Figure \ref{fig:phases_diagram}b].\\ 
\textbf{Simplification Phase 2: }To derive the forward-directed paths the students frequently followed to reach the solution they finally submitted, we removed the backward edits representing broken objectives [Figure \ref{fig:phases_diagram}c].\\
\textbf{Simplification Phase 3: }We removed edges that represent back and forth transitions corresponding to scenarios when the students brought back a bunch of codes that were removed in the immediately previous step [Figure \ref{fig:phases_diagram}d]. For example, a student reached from state S to state 34 using the path S$\Rightarrow$3$\Rightarrow$34. At state 34, the student removed lines of code that broke obj 3 and 4 and caused the transition 34$\Rightarrow$S. When the student brought back those codes, they directly moved from state S to State 34. The entire sequence of transitions is S$\Rightarrow$3$\Rightarrow$34$\Rightarrow$S$\Rightarrow$34 which was simplified to S$\Rightarrow$3$\Rightarrow$34.\\
Using the above procedure, we generated two state transition diagrams based on objective detections by the system and experts. Here, the experts are authors of this paper who constructed and verified the diagrams through agreement.\\
From the expert and system detection based simplified state transition diagrams [Figure \ref{fig:phases_diagram}d and \ref{fig:sys_detect_diagram} respectively], we extracted the frequent solution paths adopted by the students while solving Squiral. Table \ref{tab:solution_paths} lists the solution paths and respective frequencies.

\subsubsection{How DDAIF Guides Students during a Problem Solving Attempt}\label{expert_state_dg}
The frequent solution paths [Table \ref{tab:solution_paths}] extracted from the expert state transition diagram [Figure \ref{fig:phases_diagram}d] show that all frequent student solution paths end at the WC (working code) state. This represents a high task completion rate [93\%] in students using DDAIF. Solution paths of only 2 students ended at state `NWC’ [non-working code]. These paths are infrequent and are not listed in Table \ref{tab:solution_paths}. Most students closely followed the given objectives and explored corresponding code constructs. Thus, their solution paths contained predefined objective-completion based states. However, 4 students formulated solutions that indicate that they ignored the given objectives. They did not use custom blocks and wrote sequences of `move', `turn', and `increment' statements instead of implementing a loop.  Their solutions correspond to the path S$\Rightarrow$WC [Path e3 in Table \ref{tab:solution_paths}].  In these cases, none of the given objectives were detected and no feedback was given from the system. Additionally, frequent paths e1, e2, e4, and e5 in Table \ref{tab:solution_paths} indicate that students did not or could not follow the given order of objectives while working towards a solution. They often started with implementing objectives 3 and 4 on stage and moved the codes within a custom block later to complete objective 1. Note that Squiral can also be implemented by completing objectives 2, 3, and 4 on stage instead of implementing them within a custom block. However, using a custom block as indicated by objective 1 leads to a cleaner implementation and is a good programming practice (similar to using functions instead of writing long sequences of codes). This indicates that a high-quality objective list can also motivate students to adopt good programming practices. Additionally, the expert detection based state transition diagram after phase 1 simplification (\ref{fig:phases_diagram}-b) demonstrates an important visual pattern. In the diagram, the red edges are always paired with almost equally weighted black edges. These red-black edge pairs represent that the students most of the time fix a broken objective immediately by bringing back the same features which when removed caused the broken objective. This indicates that DDAIF system has the potential to lead students in the right direction when they make a mistake. 

\subsubsection{When DDAIF Failed to Guide}\label{diff_expert}
As we compared the solution paths extracted from system and expert state transition diagrams [Table \ref{tab:solution_paths}], we noticed that most of the expert detected solution paths have the prefix `S$\Rightarrow$3’, i.e. students completed objective 3 first. However, the system detected solution paths mostly contain `S$\Rightarrow$1’, i.e. students first completed objective 1. This scenario occurred because the system does not detect Objective 3 as complete unless it is implemented within a custom block (Objective 1). However, often students completed Objective 3 and sometimes the entire solution of Squiral in stage and later moved the code within a custom block to complete Objective 1. This indicates, that unlike human tutors DDAIF cannot evaluate objective completion in any order, specifically when the objectives are inter-dependent. Another noticeable difference is that in most of the system detected frequent solution paths Objective 2 was detected. However, according to experts, often students' implementation of Objective 2 was incomplete due to incorrect loop parameters [details described in Section \ref{FP_2}]. Since the DDAIF system did not extract loop parameters as frequent features required for objective 2 and only looked for the presence of nested loops, objective 2 got incorrectly detected with high frequency by the system. 

\subsubsection{Findings:} Our collective observation from the analyses to address evaluation criterion 2 suggest: 1) Although being fallible, DDAIF can guide students during programming tasks similar to a human tutor to a great extent. It provided students objectives to achieve, prevented them from going down a wrong path (by marking broken objectives as red), and kept them motivated to achieve the optimal solution (by getting all objectives green) or to adopt good programming practices, 2) Whereas a human expert/tutor can identify objectives independently, the system could require ordered completion of objectives for accurate detection specifically in case of inter-dependent objectives, and 3) Since, data-driven systems like DDAIF looks for high-level objectives constructed from frequent code features extracted from previous students' solution, it may not be able to give feedback on fine-grain infrequent code details, for example, feedback on exact loop parameter in this case.     

\begin{table}[h!]
\centering
\caption{Frequency of expert and system detection based solution paths..}
\label{tab:solution_paths}
\begin{tabular}{lll}
\hline
Path ID &                                                                                 & Frequency \\ \hline
        & \textbf{Expert Detection Based Paths}                                           &           \\ \hline
e1      & S$\Rightarrow$3$\Rightarrow$13$\Rightarrow$134$\Rightarrow$1234$\Rightarrow$WC  & 4         \\ \hline
e2      & S$\Rightarrow$3$\Rightarrow$34$\Rightarrow$134$\Rightarrow$1234$\Rightarrow$WC  & 4         \\ \hline
e3      & S$\Rightarrow$WC                                                                & 4         \\ \hline
e4      & S$\Rightarrow$3$\Rightarrow$13$\Rightarrow$134$\Rightarrow$WC                   & 3         \\ \hline
e5      & S$\Rightarrow$3$\Rightarrow$34$\Rightarrow$134$\Rightarrow$WC                   & 5         \\ \hline
e5      & S$\Rightarrow$1$\Rightarrow$134$\Rightarrow$WC                                  & 5         \\ \hline
        & \textbf{System Detection Based Paths}                                           &           \\ \hline
s1      & S$\Rightarrow$1$\Rightarrow$12$\Rightarrow$123$\Rightarrow$END                  & 4         \\ \hline
s2      & S$\Rightarrow$1$\Rightarrow$13$\Rightarrow$123$\Rightarrow$END                  & 2         \\ \hline
s3      & S$\Rightarrow$1$\Rightarrow$13$\Rightarrow$123$\Rightarrow$1234$\Rightarrow$END & 3         \\ \hline
s4      & S$\Rightarrow$1$\Rightarrow$13$\Rightarrow$134$\Rightarrow$END                  & 2         \\ \hline
s5      & S$\Rightarrow$1$\Rightarrow$13$\Rightarrow$134$\Rightarrow$1234$\Rightarrow$END & 3         \\ \hline
s6      & S$\Rightarrow$1$\Rightarrow$14$\Rightarrow$134$\Rightarrow$1234$\Rightarrow$END & 4         \\ \hline
s7      & S$\Rightarrow$1$\Rightarrow$END                                                 & 7         \\ \hline
\end{tabular}
\end{table}

\section{Evaluation Criterion 3: Effective Feedback Components}\label{how_helps}
Efficient completion of programming tasks has a direct impact on metrics to measure learning or performance (for example, test scores). We observed that 25 out of 27 students participating in this study had a working solution to the Squiral assignment, although the solutions may not be perfect [details in Section \ref{qual_fallibility}]. In this evaluation criterion, we investigate which component of our feedback system helped students to successfully complete the Squiral assignment and how. In any feedback system, these components can be called \textit{Effective Feedback Components} since they may directly impact students' programming experience, performance, and eventually learning. 

We adopted a case-based approach to address this evaluation criterion where we analyzed the cases of students Jade and Lime. These two students failed to implement Squiral when DDAIF was not available. However, when DDAIF was made available they successfully completed the task. We examined these students' sequential code traces and prepared case studies that we present next to demonstrate how the feedback system helped the students to fill the gaps in their code and led them to working solutions.
\subsubsection{Case Study Lime}
\begin{figure}
\caption{Student Lime's Solutions.}
\label{fig:x}
\begin{center}
\subfloat[Student Lime's Solution when no Feedback was given.]{
\includegraphics[width=.45\linewidth]{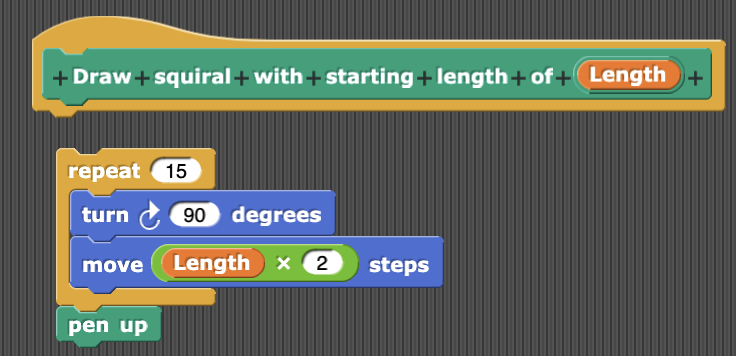}
}\quad
\subfloat[Student Lime's Solution when Feedback was Given.]{
\includegraphics[width=.45\linewidth]{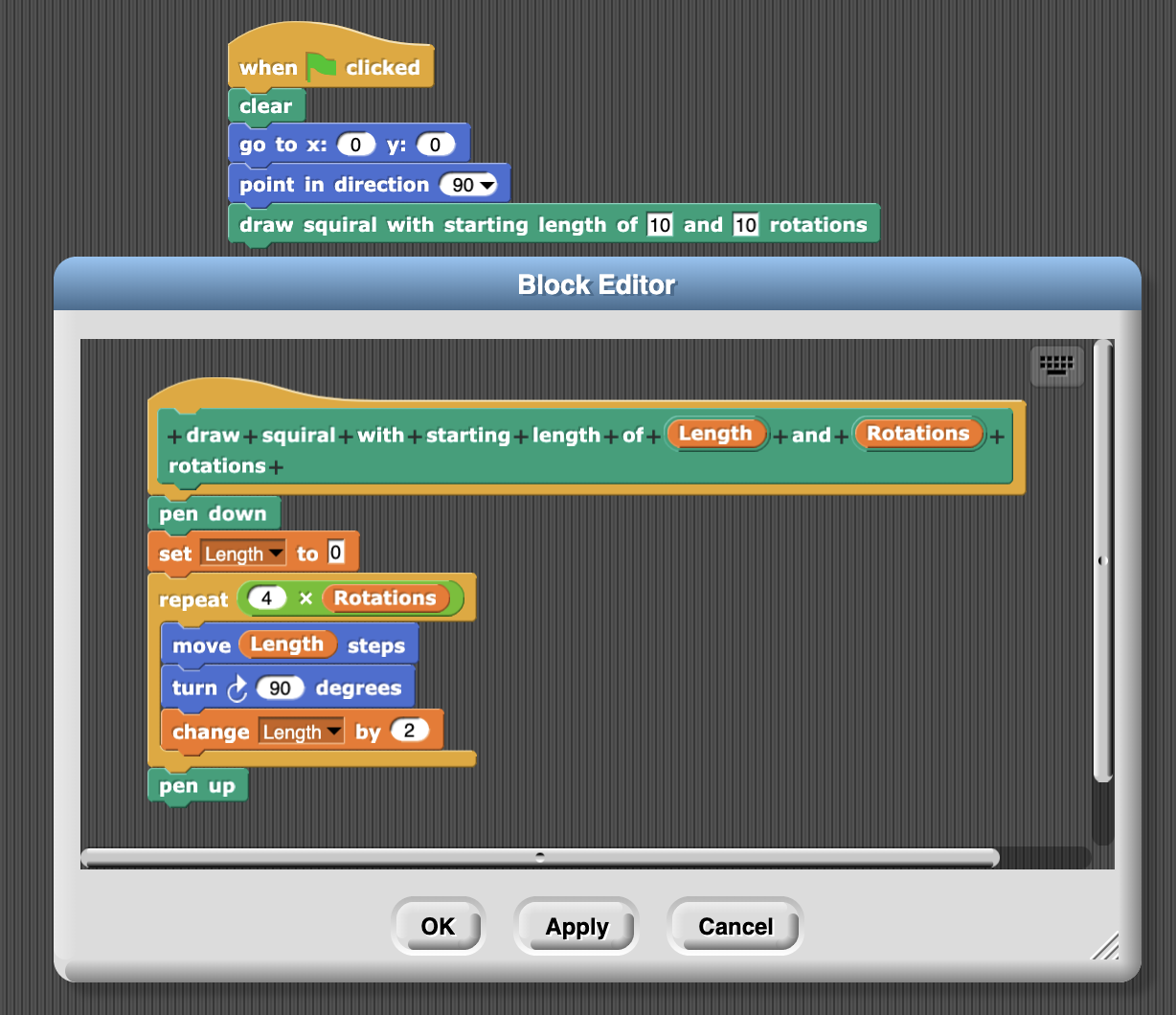}
}
\end{center}
\end{figure}

\textbf{Student, Lime, without DDAIF: }Student Lime, when attempting to solve Squiral without any hints or feedback [Figure \ref{fig:x}-a] given, used a custom block with one parameter. The student used `move' and `turn' statements within a loop in the custom block. However, there were three gaps in the code that the student could not figure out. First, a nested loop was required to iterate for `rotation count x 4' times. Second, the move statement used `length x 2' as its parameter whereas only `length' would be sufficient. Finally, the variable used in the `move' statement must be incremented at each iteration. The student spent 18 minutes and 18 seconds before giving up, being unable to figure out these issues.

\textbf{Student Lime with DDAIF: }
Later, student Lime attempted the homework again after receiving notice that the DDAIF system was made available. When Lime received feedback, they figured out the 3 issues and reached a correct solution [Figure \ref{fig:x}-b]. The second objective suggests that there is a correct number of rotations that are needed to be used within the custom block. With this feedback, Lime used `4 x Rotations' in the `repeat' block instead of using `15' and completed the second objective. The third objective suggests the use of a variable in the `move' statement. Lime used an initialized variable `length' in the `move' statement instead of `length x 2' and the objective was marked green. Finally, Lime incremented `length' within the loop and all objectives were completed and they reached a correct solution. With DDAIF, Lime spent 29 minutes 51 seconds before reaching the correct solution. Recall that Lime gave up with an incorrect solution after around 18 minutes when no feedback was given.

\subsubsection{Case Study Jade}
\begin{figure}
\caption{Student Jade's Solutions.}
\label{fig:y}
\begin{center}
\subfloat[Student Jade's Solution when no Feedback was given.]{
\includegraphics[width=.45\linewidth]{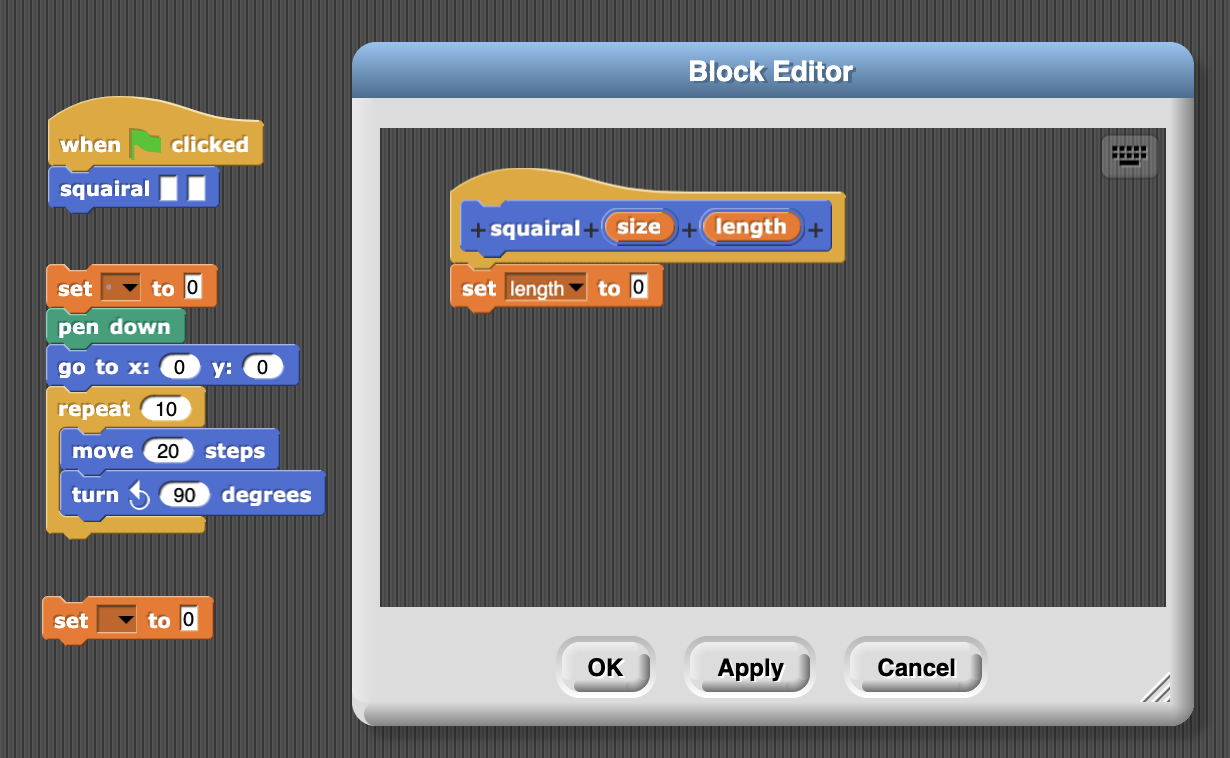}
}\quad
\subfloat[Student Jade's Solution when Feedback was Given.]{
\includegraphics[width=.45\linewidth]{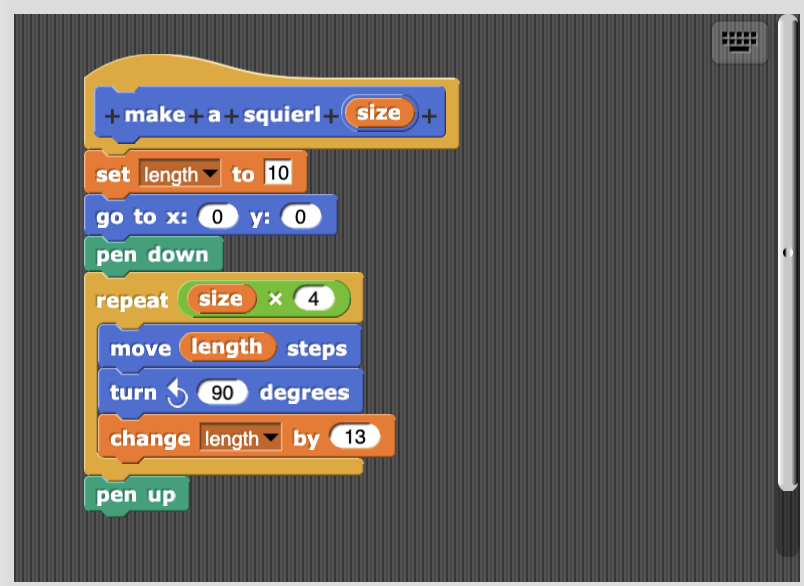}
}
\end{center}
\end{figure}

\textbf{Student Jade without DDAIF: }Student Jade initially attempted to solve Squiral without data-driven positive feedback, spent 16 minutes and 55 seconds before giving up with an incorrect solution. Jade's code [Figure \ref{fig:y}-a] contains `repeat', `move', and `turn' statements on the stage. Jade created a custom block and only used the block to  initialize a variable, `length', that was also a parameter to the block. The components to complete the objectives were partially there in Jade's code but it suffered from organizational issues. Also, Jade couldn't figure out that the `move' statement should use a variable instead of a constant and the same variable needs to be incremented at each iteration. The number of repetitions in the repeat block was also incorrect.

\textbf{Student Jade with DDAIF: }
Like Lime, Jade attempted the homework again when the DDAIF system was provided. When given feedback, Jade first created a custom block and used it on the stage which got the first objective marked green. The second objective hints at using a loop that repeats for the correct number of rotations within the custom block. This time Jade implemented the loop within the block and got the second objective correct. Within the loop, Jade used `move', `turn', and `change' statements and reached the correct solution [Figure \ref{fig:y}-b] with all objectives marked green. With adaptive feedback, it took Jade 14 minutes and 29 seconds to reach a correct solution. Whereas without feedback, Jade gave up with an incorrect solution after spending over 16 minutes on the problem.

\subsubsection{Findings}
The 2 case studies of Lime and Jade demonstrated that DDAIF was able to help students in filling up the gaps in their code to reach a correct solution. The objective list gave them smaller and easier subgoals to work on (i.e. reduced difficulty) and the completion-based adaptive immediate feedback continuously assessed their work which kept them on track. In one case, this achievement came at the cost of a higher active time (which also indicates DDAIF can improve persistence and motivation) and in the other case, the student reached a correct solution in less time (i.e. improved efficiency) when feedback was provided.
\section{Evaluation Criterion 4: Impact of Fallibility}\label{when_wrong}
Towards this evaluation criterion, we conducted two analyses: 1) A quantitative analysis focusing on what types of negative impact could occur due to system fallibility and how often; and 2) A qualitative case-study based approach that demonstrated how different impacts occurred.
\subsection{Quantitative Impact Analysis of System Fallibility}\label{impact_analysis}
For the quantitative analysis of the impact of the fallibility of DDAIF, first, we tagged the first time detection of each objective (since students are likely to be impacted more by the quality of first time detection) with one of the types [Correct (CD/CND), Incorrect (ID/IND), Early (E), or Late (L)] described in Table \ref{tab:detTypes}. Then, we categorized and explained the observed negative impacts on students' code traces [Section \ref{type_of_impact}]. Finally, to quantify the association between different types of faulty feedback (ID/IND, E, and L) and negative impacts, we documented co-occurrences of faulty feedback and negative impacts. We report the findings of our impact analysis in Section \ref{findings_6_1}. 
\subsubsection{Types of Negative Impacts}\label{type_of_impact}
\begin{figure}
\caption{State transition diagrams for a student whose solution path was impacted (IPB) due to inaccurate system detections.}
\label{fig:STD_IPB}
\begin{center}
\subfloat[System state transition diagram.]      {
\includegraphics[height=1.5in,width=.3\linewidth]{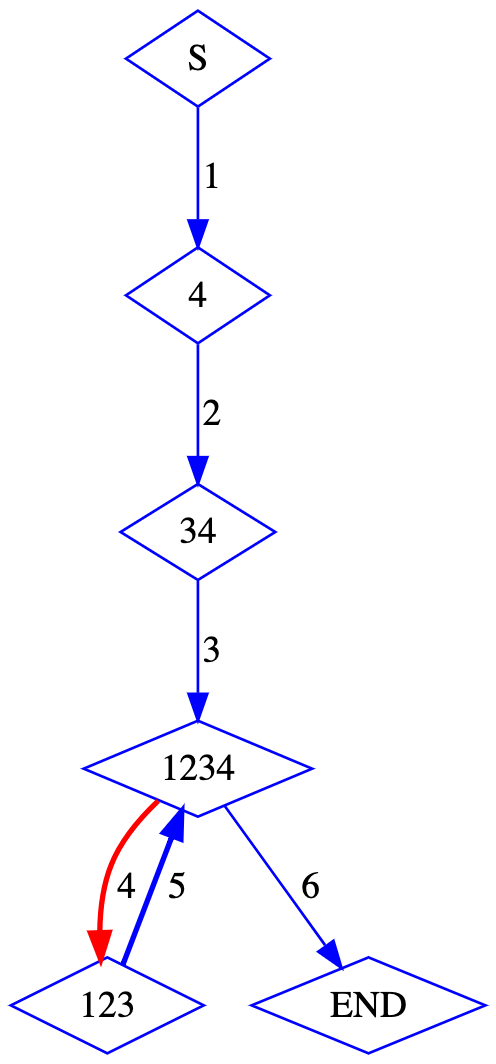}
}\quad
\subfloat[Expert state transition diagram.]      {
\includegraphics[width=.45\linewidth]{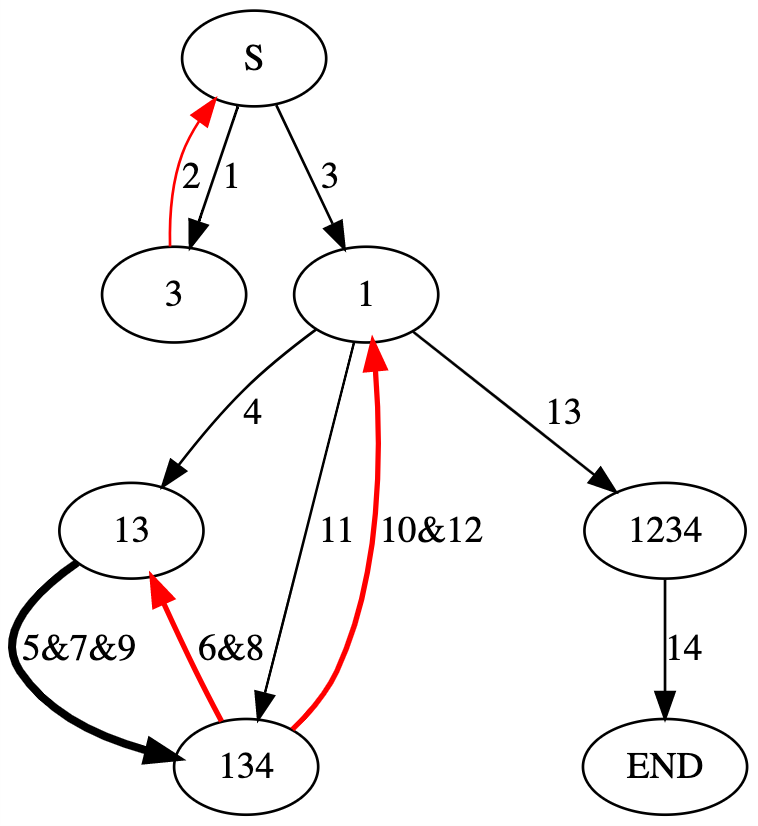}
}
\end{center}
\end{figure}

\begin{figure}
\caption{State transition diagrams for a student who spent extra time due to inaccurate system detections.}
\label{fig:STD_ITS}
\begin{center}
\subfloat[Expert diagram with 5 mins. data.]      {
\includegraphics[height=1.2in,width=.15\linewidth]{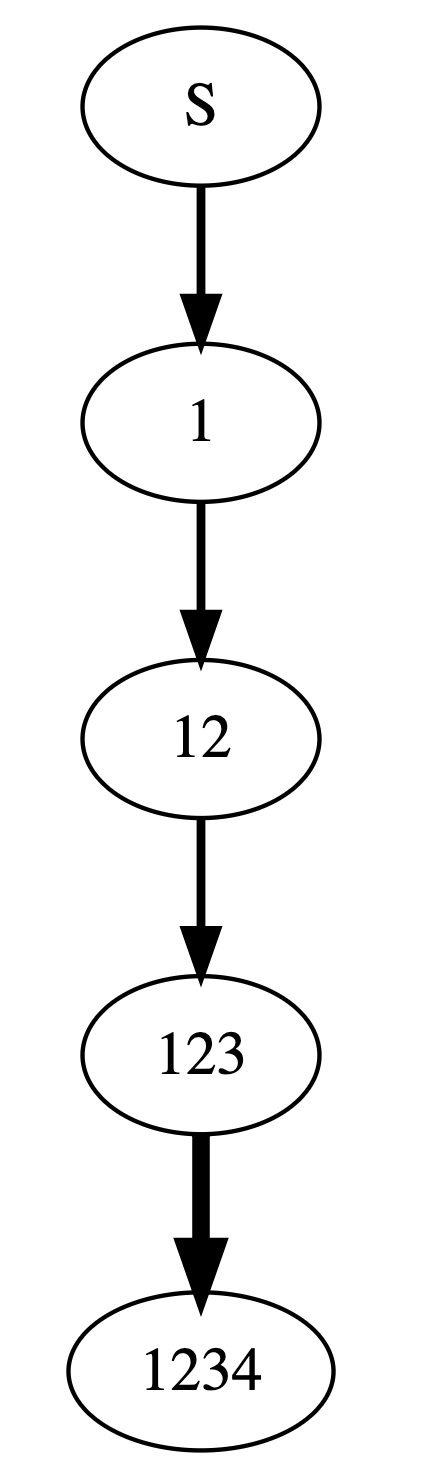}
}\quad
\subfloat[Expert state transition diagram at the end of the attempt.]{
\includegraphics[width=.3\linewidth]{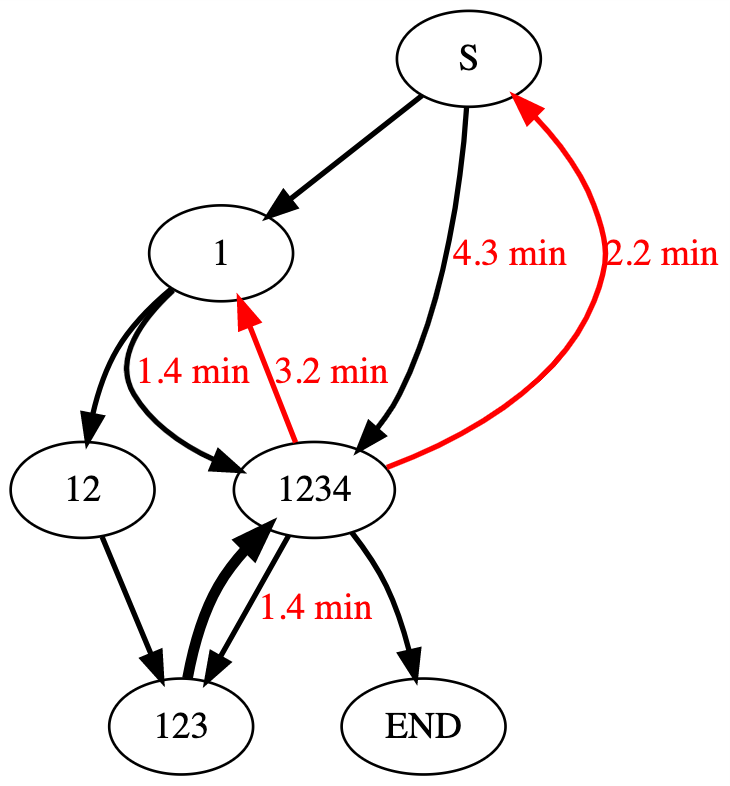}}\quad
\subfloat[System state transition diagram.]{
\includegraphics[height=1.2in,width=.3\linewidth]{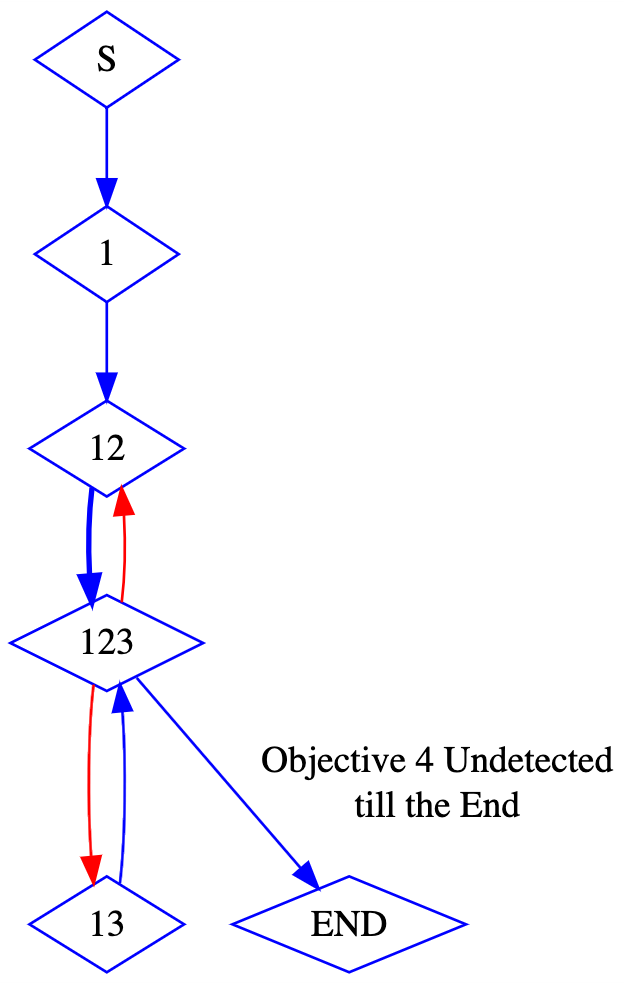}
}
\end{center}
\end{figure}

\begin{figure}
\caption{State transition diagrams for a student who early submitted a partially correct solution due to inaccurate system detections.}
\label{fig:STD_ES}
\begin{center}
\subfloat[System state transition diagram.]{
\includegraphics[height=1.5in,width=.20\linewidth]{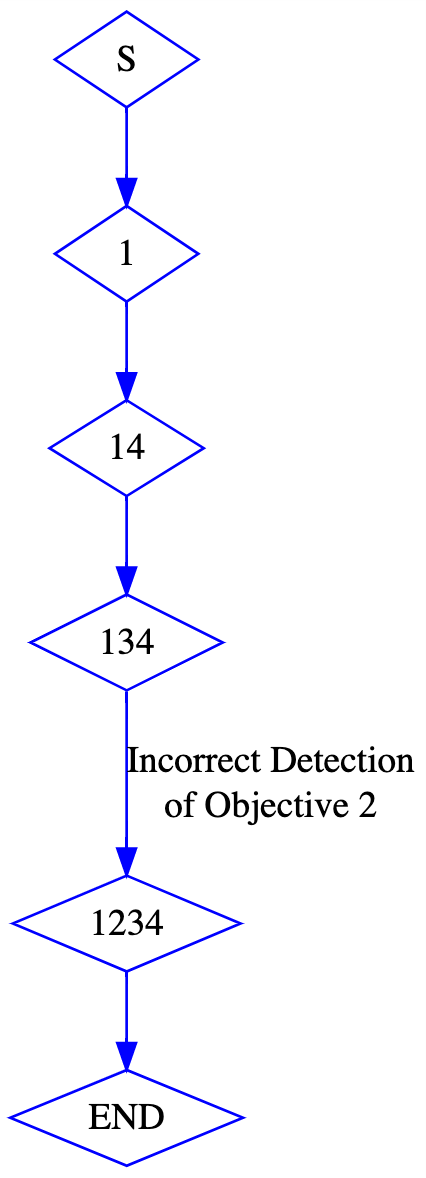}
}\quad
\subfloat[Expert state transition diagram.]{
\includegraphics[height=1.5in,width=.20\linewidth]{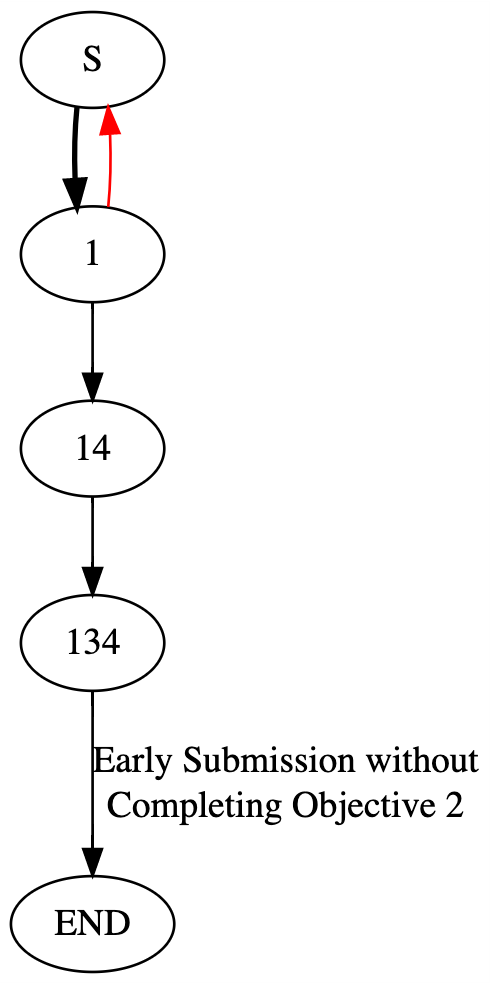}
}
\end{center}
\end{figure}

The expected impacts of the DDAIF system on student outcomes are threefold: 1) \textit{Optimality} - pushing the students towards an optimal solution through the completion of objectives; 2) \textit{Efficiency} - decreased active and idle time as indications of increased efficiency and motivation; and 3) \textit{Accuracy} - submission of a correct solution to a programming problem. Any observed impact that goes against these 3 expected impacts is considered a negative impact. We formalize and explain different types of negative impacts that we observed in students' code traces below:\\
\textbf{\textit{Type 1: Impact on programming behavior (IPB): }}IPB refers to the situation when following the feedback given by the DDAIF system influences a student to keep code constructs that are not optimal, incorrect, or only partially correct.\\
\textit{Example Scenario: A student did not use any custom block in their code. However, objective 1 was detected incorrectly (ID) by the system. The student did not use a custom block in their solution till the end, although it was the optimal approach.}\\
Figure \ref{fig:STD_IPB} further explains the type. It presents the system and expert detection-based state transition diagrams for a student who got 3 early objective detections (E cases) and 1 correct detection (CD case) while solving Squiral. According to system detections, the student's solution path was S$\Rightarrow$4$\Rightarrow$34$\Rightarrow$1234$\Rightarrow$123$\Rightarrow$1234$\Rightarrow$END, indicating the student reached state `1234’ twice. However, the first time when the student reached state `1234’ according to the system, none of the objectives were completed according to experts. But, the student continued following system feedback and kept incorrect codes till the point of ending up at a non-working implementation with all 4 objectives marked green. It was only then that the student possibly realized the fallibility of the system. The student was later observed to explore different solution paths ignoring the feedback to actually reach the `1234’ state(Figure \ref{fig:STD_IPB}-b).\\
\textbf{\textit{Type 2: Impact on time spent (ITS): }}ITS refers to the scenario when the DDAIF system impacts the time a student spends in the system.\\
\textit{Example Scenario 1: ID or E detections led a student to a non-working solution and the student had to spend more time to get back on the right track (lingering impact of IPB).\\
Example Scenario 2: The student completed an objective, but the system did not detect it (IND/L). The student spent extra time on that objective thinking that their solution is not correct or to make the system detect that objective.}\\
To visualize ITS, we generated state transition diagrams (Figure \ref{fig:STD_ITS}) for a student who completed objective 4 which the system never detected [Figure \ref{fig:STD_ITS}-c)]. The student reached a correct solution as depicted in Figure \ref{fig:STD_ITS}-a within 5 minutes of starting the attempt. The student spent 12.5 minutes more making unnecessary changes [ Figure \ref{fig:STD_ITS}-b shows spent time along edges] before submitting the program.\\
\textbf{\textit{Type 3: Early Submission (ES): }}ES refers to the situation in which relying on the feedback given by the DDAIF system leads a student to submit a partially correct solution.\\
\textit{Example Scenario: ID or E detection made the student think that their solution is    
complete and they submitted when their solution was only partially correct (lingering impact of IPB).}\\
Figure \ref{fig:STD_ES} presents the system and expert state transition diagrams for a student who submitted a partially correct version of the Squiral program after objective 2 was early detected by the system (Figure \ref{fig:STD_ES}-a). However, the expert state transition diagram (Figure \ref{fig:STD_ES}-b) indicates objective 2 was incomplete.
\subsubsection{Findings}\label{findings_6_1}
Among a total of 108 cases of first-time objective detection feedback (27 students x 4 objectives), 26 cases (24\%) were completely incorrect ( ID, and IND cases in Column 1, Table \ref{tab:detTypeImpact}) and 38 cases (35\%) are partially incorrect (E, and L cases in Column 1, Table \ref{tab:detTypeImpact}). Among a total of 64 cases of faulty ID, IND, E, and L feedback, 26 feedback (40\%) co-occurred with at least one of the defined negative impacts [details shown in Table \ref{tab:detTypeImpact}]. We observed incorrect detection (ID cases) to have the highest impact ratio (\textit{78.57\%} ID cases co-occurred with a negative impact). These ID cases were observed to lead students to all 3 types of negative impact (IPB, ITS, and ES) [Table \ref{ref:impactCauseCount}]. \textit{25\%} IND detections [Table \ref{tab:detTypeImpact}] were associated with spending more time on completed objectives (ITS) [Table \ref{ref:impactCauseCount}]. E cases co-occurred with increased active time (ITS) [Table \ref{ref:impactCauseCount}] spent to fix an early detected objective that potentially made the students keep incorrect/partially correct code features until the code failed to produce the expected outcome. On the other hand, L cases co-occurred with scenarios when students tried out multiple ways to complete the same objective (IPB) possibly thinking that their solution was incorrect or to have all the objectives detected by the DDAIF system before submission resulting in increased time (ITS) [Table \ref{ref:impactCauseCount}]. Note that often multiple detections were conjointly responsible for a single case of impact.
\begin{table}[ht!]
\centering
\caption{Types of First Time Detection of Objectives.}
\label{tab:detTypes}
\begin{tabular}{ll}
\hline
Detection Type        & Explanation                                                                                                                                                                 \\ \hline
Correct (CD or CND)   & \begin{tabular}[c]{@{}l@{}}The system detected a completed \\ objective on-time (CD) or the system\\ did not mark an incomplete objective\\ as complete (CND).\end{tabular} \\ \hline
Incorrect (ID or IND) & \begin{tabular}[c]{@{}l@{}}The system detected an incomplete\\ objective (ID) or the system did not \\ detect a completed objective at all\\ (IND).\end{tabular}            \\ \hline
Early (E)             & \begin{tabular}[c]{@{}l@{}}The system detected an objective slightly\\ before the student actually completed\\ it.\end{tabular}                                                      \\ \hline
Late (L)              & \begin{tabular}[c]{@{}l@{}}The system detected an objective after\\ sometime the student actually completed\\ it.\end{tabular}                                              \\ \hline
\end{tabular}
\end{table}
\begin{table}[h!]
\centering
\caption{Ratio of different types of detections having unintended impacts.}
\label{tab:detTypeImpact}
\begin{tabular}{lllllll}
\hline
\begin{tabular}[c]{@{}l@{}}Det. Type \\ and Count\end{tabular} & Obj1 & Obj2 & Obj3 & Obj4 & Impacted & Ratio   \\ \hline
CD (22)                                                        & 9    & 0    & 11   & 2    & -        & -       \\ \hline
CND (22)                                                       & 1    & 11   & 5    & 5    & -        & -       \\ \hline
ID (14)                                                        & 5    & 9    & 0    & 0    & 11       & 78.57\% \\ \hline
IND (12)                                                       & 0    & 0    & 5    & 7    & 3        & 25\%    \\ \hline
E (29)                                                         & 11   & 5    & 2    & 11   & 9        & 31\%    \\ \hline
L (9)                                                          & 1    & 2    & 4    & 2    & 3        & 33.33\% \\ \hline
\end{tabular}
\end{table}
\begin{table}[h!]
\centering
\caption{Count of co-occurrences of different types of impacts and detection types.}
\label{ref:impactCauseCount}
\begin{tabular}{lll}
\hline
Impact Type  & Co-occurrence Count & \begin{tabular}[c]{@{}l@{}}Detections Co-occurring\\ with the Impact\end{tabular} \\ \hline
Type 1 (IPB) & 3                   & ID, L                                                                             \\ \hline
Type 2 (ITS) & 7                   & ID, IND, E, L                                                                     \\ \hline
Type 3 (ES)  & 7                   & ID                                                                                \\ \hline
\end{tabular}
\end{table}
\subsection{Qualitative Impact Analysis of System Fallibility: Case Studies}\label{qual_fallibility}
In this section, we present three case studies demonstrating how the fallibility of DDAIF impacted students while solving Squiral. The case studies were developed by experts from observations accumulated as they replayed students' Squiral solving attempts.

\subsubsection {ID/E Detections Leading Students to an Incorrect solution}\label{FP_1}
\begin{figure}
\centering
\includegraphics[width=0.5\textwidth]{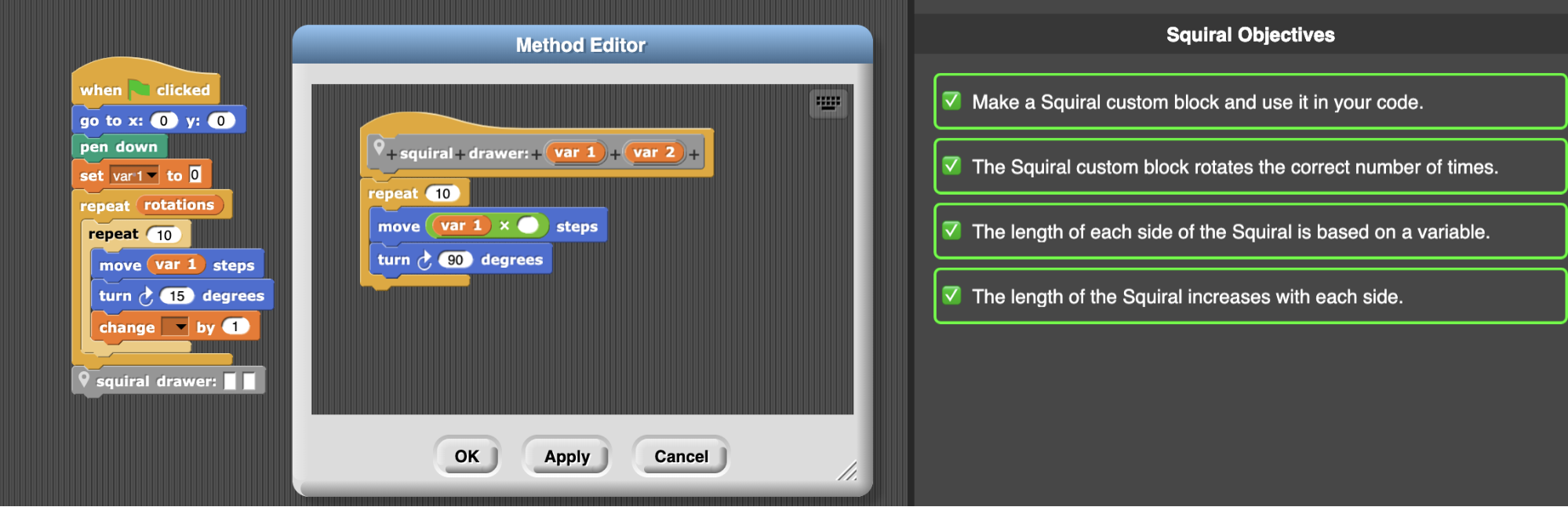}
\caption{Incorrect Solution Initially Implemented by Student Cyan.}\label{fig:C2}
\end{figure}

Here we present the scenario when the system marked an incomplete objective as completed and following this incorrect feedback led students to an incorrect solution. Depending on whether the students’ codes were capable of drawing a Squiral or not, we observed two different student behaviors - 1) when students' codes were not capable of drawing a Squiral, they spent more time and tried out other approaches till they got the correct solution (recall IPB and ITS), and 2) when students' codes were capable of drawing a Squiral, the students early submitted partially correct solution (recall ES) [depicted in Section \ref{FP_2}]. In this section, we present the case study of `Cyan’ as a representative to demonstrate the first behavior and then we discuss other similar cases to discuss the generalizability of the selected case.\\
\textbf{Case Study Cyan:}
Student Cyan created a custom block and used it on the stage and got the first objective correct. Cyan used two parameters in the custom block and used one of them in a `move' statement within a nested `repeat' block that got him the second and third objectives correct. However, Cyan implemented another nested loop and added a `change' statement within that loop in the stage instead of adding them to the custom block. The system detected the objective  and marked the fourth objective green. At this point, the student Cyan had all objectives correct but the code [Figure \ref{fig:C2}] was unable to draw a Squiral.

Later, Cyan removed the `change' statement from the stage causing the fourth objective to be broken. However, removing the custom block from the stage was causing other objectives to be broken. Cyan then moved the `change' statement to the custom block and corrected the rotation count in the `repeat' statement. At this point, the solution was correct and similar to Figure \ref{fig:expert_solution_output}a. The incorrectly detected objectives led Cyan to a non-working solution. In this case, Cyan had to ignore the detectors and do extra work to reach a correct solution.

\textbf{Generalizability of Selected Case:} Three other students had a similar situation. One of them had a working program that was drawing three sides of the Squiral using an inner loop and one side manually. Although this implementation was not considered fully correct by the researchers, the four objectives were detected by the system and the student submitted the partially correct solution. The other two students had a non-working program with syntactical problems when the system detected all four objectives. Incorrect/missing output eventually compelled each student to modify their codes ignoring the system's feedback and to reach a 100\% correct solution at the end.

\subsubsection{IND/L Detections Causing Students to Work More than Necessary}
\begin{figure}
\centering
\includegraphics[width=0.5\textwidth]{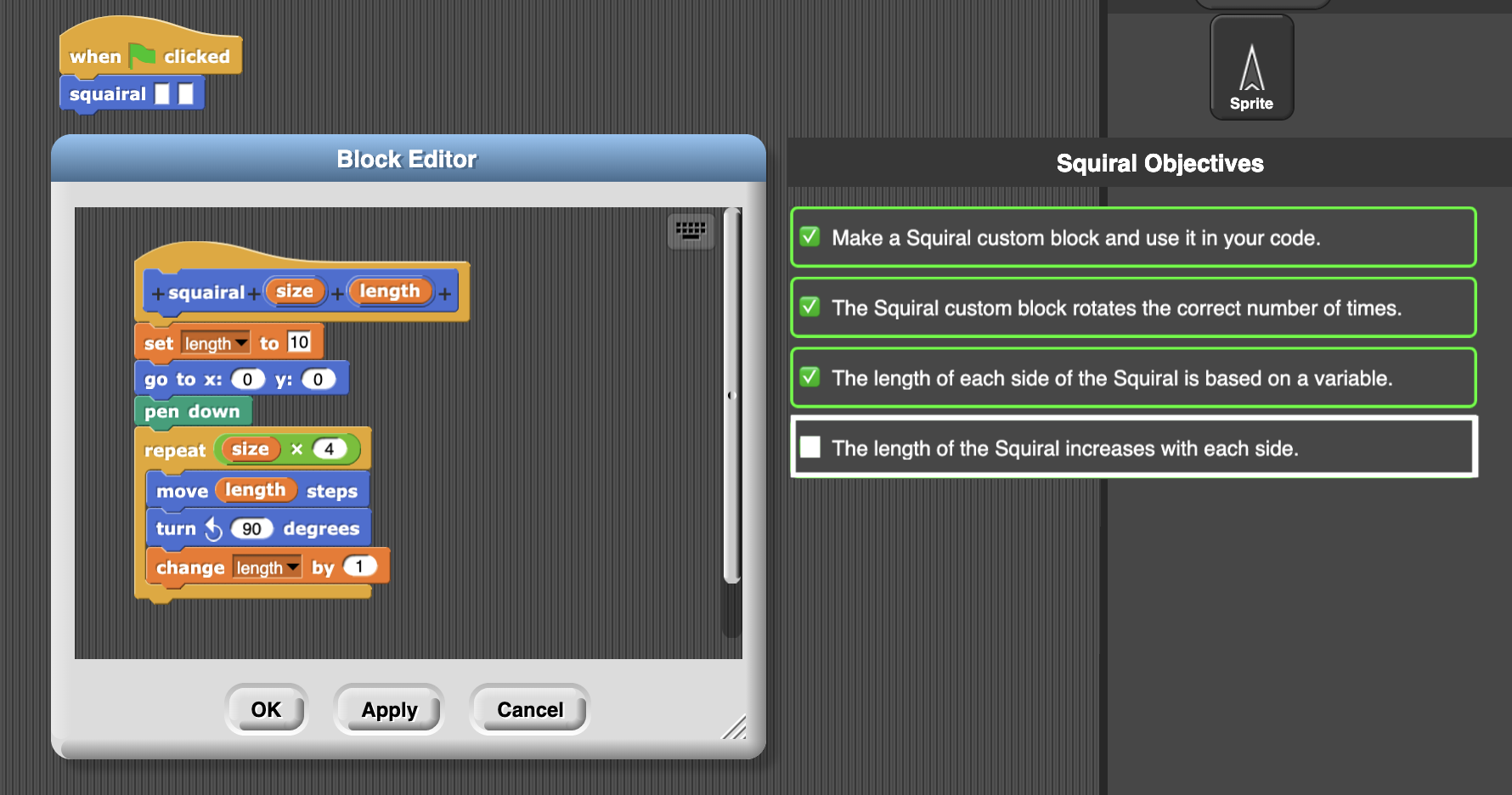}
\caption{Correct Solution Initially Implemented by Azure.}\label{fig:c1}
\end{figure}

Here we present the case study of student `Azure’ demonstrating the scenario when the DDAIF system could not detect completed objectives potentially causing students to think that their solution is incorrect leading to unnecessary work (IPB) and increased active time (ITS). Later we briefly discussed the case of student `Blue’ where we observed a similar scenario to demonstrate the generalizability of our selected case.\\
\textbf{Case Study Azure:}
Student Azure started solving Squiral by creating a custom block and got the first objective correct. Azure used 2 parameters, `size' and `length', to denote the number of rotations and length of the first side of the innermost loop. They created a loop with a `repeat' block  with the correct number of rotations (`size x 4') and got the second objective correct. As Azure used the `length' parameter in the `move' statement within the loop, they got the third objective correct. Then Azure added a `turn' statement and incremented the `length' variable. At this point, the fourth objective was completed, according to researchers. However, the objective was undetected by the system, because Azure used a `turn' statement that was different from those used in the previous students’ solutions [that were used to extract and detect the objectives]. According to researchers, Azure's solution was 100\% correct at this point [Figure \ref{fig:c1}]. It took this student only 2 minutes and 24 seconds to reach the correct solution.

However, the fourth objective was not detected by the system. Azure kept working on their code. Azure made several changes to their code which led them to an incorrect solution. Finally, Azure ended up submitting a solution that was also 100\% correct according to researchers but was slightly different from their initial  solution. In the submitted solution, Azure removed the `length' variable from the parameter list of the custom block. The fourth objective was still undetected. While doing these changes, the student spent 12 minutes 5 seconds more in the system which is almost 5 times the amount of time the student spent to get a correct solution in the first place.

\textbf{Generalizability of Selected Case:} We observed a similar situation in the case of student Blue who reached a correct solution at around 1 hour. But objective 4 was undetected by the system. Blue kept working for another 36 minutes(almost 50\% of the time taken to reach the correct solution at the first attempt).

\subsubsection{ID/E Detections Causing Students to Stop Early at a Partially Correct Solution}\label{FP_2}
\begin{figure}
\centering
\includegraphics[width=0.5\textwidth]{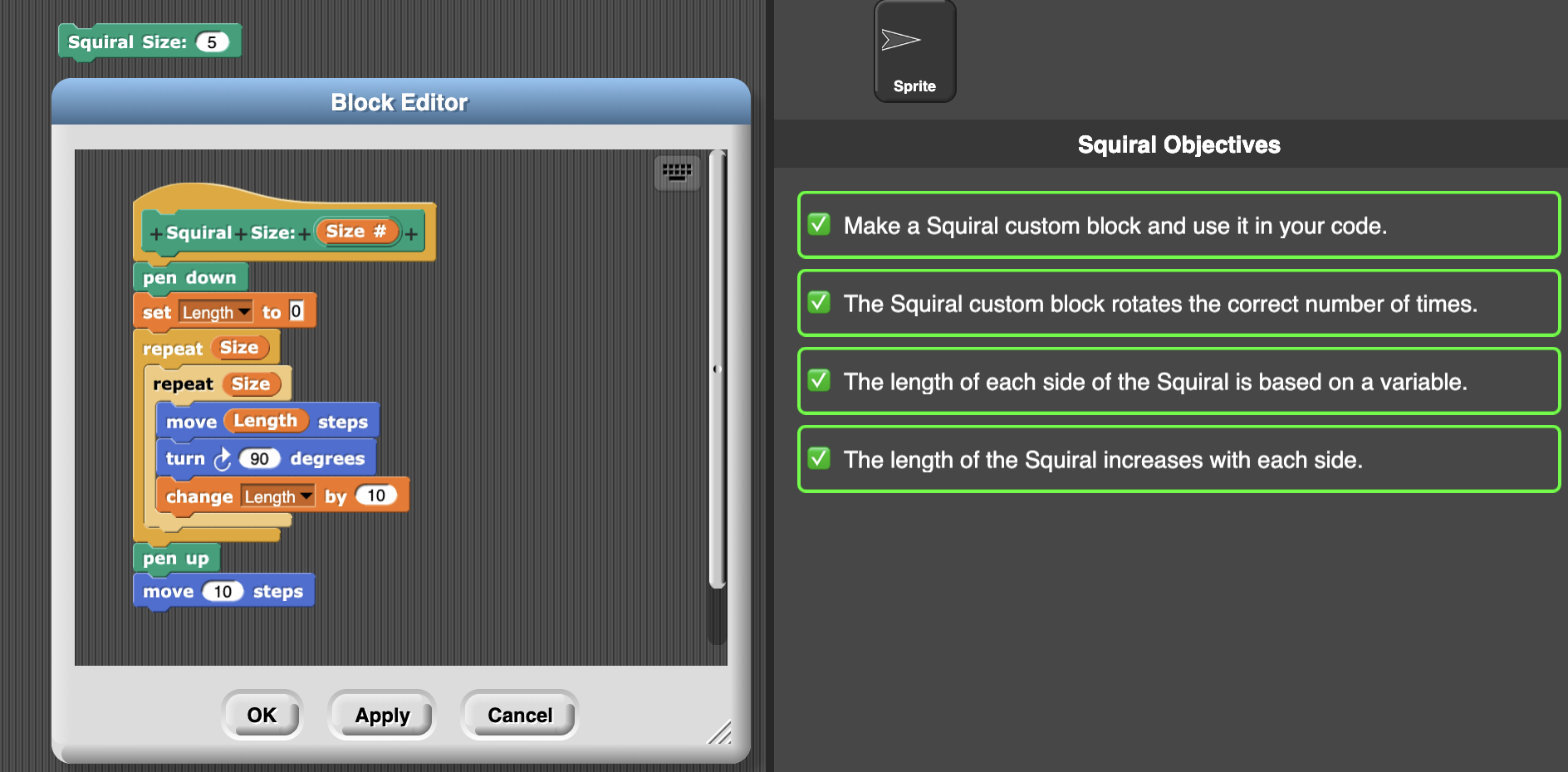}
\caption{Solution Submitted by Student Indigo}
\label{fig:C3}
\end{figure}

In this section, we present a case study (Case Study Indigo) to represent the scenario when a ID/E case caused students to consider a partially correct solution as correct and stop their attempt early (recall ES). Later, we discussed 5 other similar cases to establish the generalizability of the selected case.\\
\textbf{Case Study Indigo: }
Student Indigo’s solution [Figure \ref{fig:C3}] had objectives 1, 3, and 4 correctly completed according to researchers and the objectives were detected by the objective detection system as well. Indigo created a custom block and used it in the stage [required to complete objective 1]. They used `pen down' and added `move', `turn', and `change' statements accordingly [required to complete objective 4] within a nested loop implemented with two `repeat' statements. In the `move' statement, Indigo used an initialized variable, `Length' [required to complete objective 3], and increment the value of `Length' by 10 at each iteration. However, the rotation count used in the nested loop was wrong. One of the `repeat' statements should have the count of rotations and the other should have a constant 4, indicating the 4 sides of the square drawn at each rotation. The objective detection system detected the use of the nested loop and marked objective 2 green. The code was able to draw a Squiral. However, the implementation was not completely correct. But, once the student Indigo got 4 objectives correct, they submitted their solution.

\textbf{Generalizability of Selected Case: }We found 5 other cases with a scenario similar to indigo. Each of them had an incorrect loop structure. They used `size x size', `size x sides' (none of the variables contained the value 4), `1 x rotations', `2 x sides',  and `5 x y' respectively which clearly indicated their logical misconception about the loop structure required to draw squares repeatedly. However, the system detected objective 2 and the students submit their code without correcting this issue.  

\subsubsection{Findings}
We observed cases where students reached a correct solution, but possibly considered their solutions incomplete and worked unnecessarily for a longer time when their completed objectives were not detected by the system. In cases of students having incorrect non-working solutions with the system having marked all 4 objectives, the students modified their codes overriding the objective detectors to reach a working solution possibly realizing that only following the system is not enough. Our third case study showed that students having a working program with programmatic problems did not self-evaluate and submitted partially-correct solutions. All the case studies indicate students' high reliance on system feedback since they did not seem to question the feedback system or self-evaluate to improve their program as long as they had a working program that drew the correct shape. However, in the cases when objectives were not detected, students again avoided self-evaluation and even ignored the produced output that showed they had a working solution and continued working. 
\section{Evaluation Criterion 5: Generalized Student Response}\label{analysis_responses}
\begin{figure}
\centering
\includegraphics[width=0.24\textwidth]{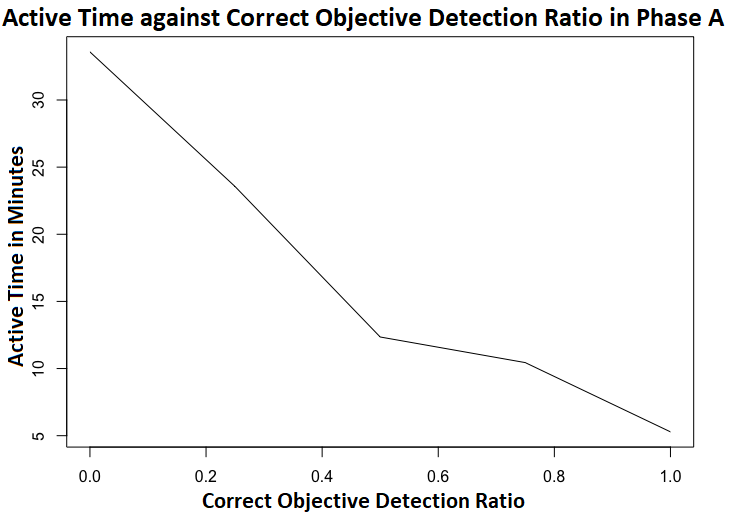}
\includegraphics[width=0.24\textwidth]{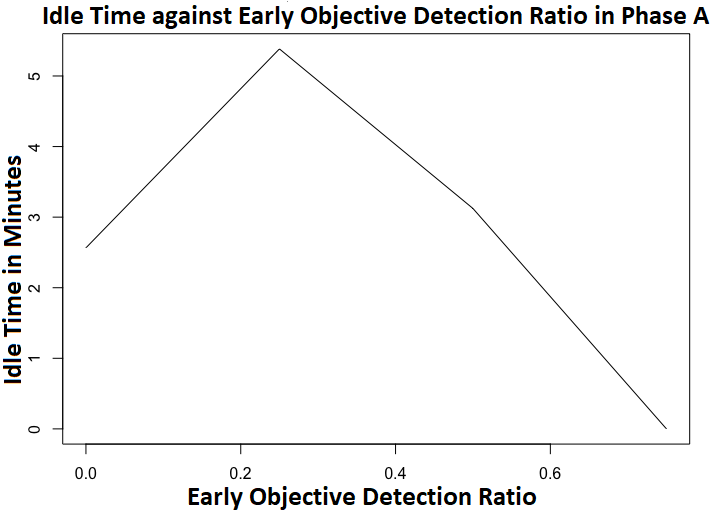}
\caption{a) Active Time in Phase A against Correct Objective Detection Ratio; b) Idle Time in Phase A against Early Objective Detection Ratio}\label{fig:phaseA}
\end{figure}

\begin{figure}
\centering
\includegraphics[width=0.24\textwidth]{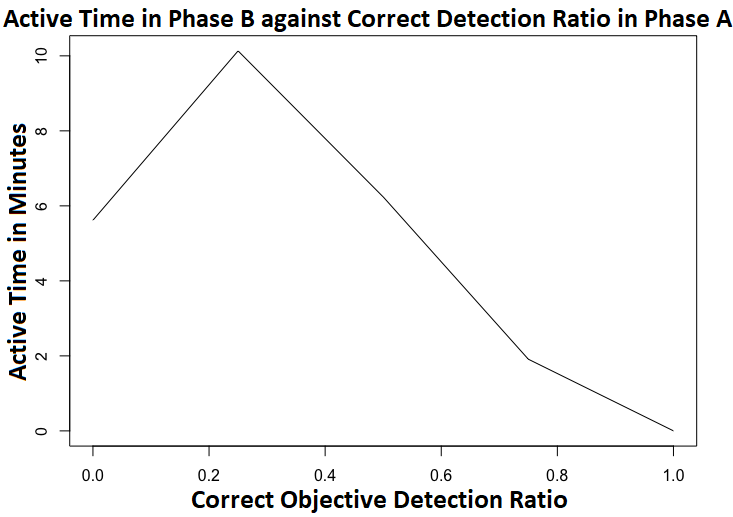}
\includegraphics[width=0.24\textwidth]{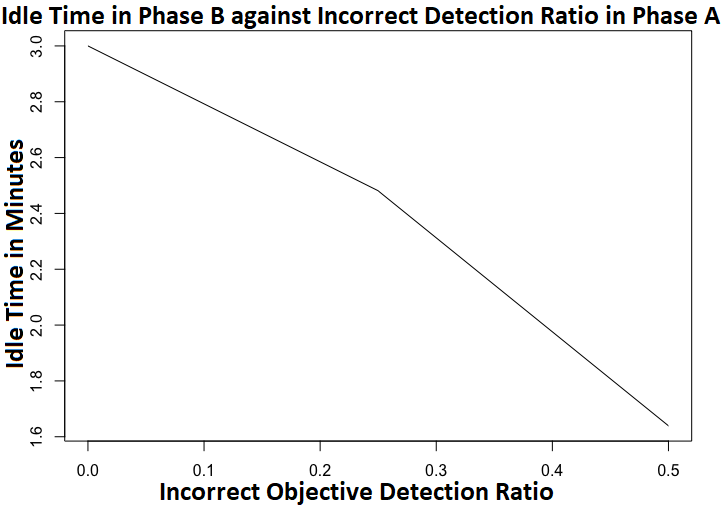}
\caption{a) Active Time in Phase B against Correct Objective Detection Ratio in Phase A; b) Idle Time in Phase B against Early Objective Detection Ratio in Phase A}\label{fig:phaseB}
\end{figure}

The purpose of this evaluation criteria is to identify patterns in behavior in response to the feedback received generalizable over the entire student population. To discover behavioral patterns, we looked into students' problem-solving attempts in different phases (higher granularity) rather than examining each edit (lower granularity) and analyzed how the feedback received in the early phases of a problem-solving attempt impacted later phases of the attempt. We divided the total time each student spent on the system into three phases: a) Phase A: when objectives were detected for the first time; b) Phase B: when changes were made to previously detected objectives. c) Phase C: when students spent time in the system without changing any objective around the end of their solution attempts. For example, a student got objectives 1, 3, and 4 marked green within 20 minutes of starting the attempt [Phase A]. Then the student spent another 10 minutes breaking and correcting previously detected objectives several times [Phase B]. Finally, the student spent another 5 minutes [Phase C] when no change in any of the objectives was detected. We tried to relate correct, incorrect, early, and late detection ratios in phase A with the active (time spent in making code edits) and idle (large time gaps $>3$ minutes in between code edits) time spent in phases A, B, and C to understand if the detection types regulate the time or effort students put on the assignment.\\
\textbf{Active and Idle Time Observed in Phase A: }
In this phase, objectives got detected for the first time by the system. Students spent wide range of time before an objective was detected (Range of active time: 1-84.5 mins., range of idle time: 0-19 mins.). We plotted average active and idle time against correct objective detection ratios and observed that students with higher correct objective detection ratios have shorter phase A in terms of active time [Figure \ref{fig:phaseA}a]. In this phase, only a few cases were found when the students had idle time. 16 out of 27 students did not have any idle time at all. 7 students had idle time ranging from 3-to 5 minutes. The rest of the 4 students had idle time ranging from 10 - 19 minutes. We observed that a higher early detection rate (over 25\%) has a decreasing trend in average idle time [Figure \ref{fig:phaseA}b]. This may potentially indicate that positive feedback can be motivating to students, even if it is provided early.\\
\textbf{Active and Idle Time Observed in Phase B: }For 11 students, phase B did not occur at all, due to either the fact that an objective was never detected, or the students immediately submitted their code after all of the objectives were detected for the first time in phase A. 10 students spent $>$0 - $<$10 minutes, and 6 students spent $>$10 - 35 minutes in phase B. 4 of the 6 students who spent a higher amount of time in this phase B had a high early detection ratio at phase A (50-75\%), and 1 student had a high incorrect detection ratio (50\%). These students did not have correct solutions, even if some or all of the objectives got detected in phase A. In this phase B, 21 out of 27 students had no idle time. 6 students had idle time ranging from 3 to 24 minutes. As we plotted average active and idle time in phase B against the correct and incorrect objective detection ratio in phase A, we observed a higher correct detection rate ($>$25\%) in phase A seemed to decrease the active time spent in phase B [Figure \ref{fig:phaseB}a]. This indicates correct objective detections in phase A pushed the students towards the end of their attempt. However, incorrect objective detection in phase A decreased idle time in phase B and caused the students to continue actively working [Figure \ref{fig:phaseB}b].\\ 
\textbf{Active and Idle Time Observed in Phase C: }
Phase C indicates the time when no change was detected in any of the objectives. In this phase C, one of the following scenarios occurs: 1) the student had a working solution with most of the objectives detected in the earlier phases and was making minor modifications without impacting the objectives; or 2) At least one of the objectives were undetected and the student was working on the assignment but submitted the attempt without another objective being detected. We observed that when the first scenario occurred for 18 out of 27 students, they spent only 0.1-8 minutes in phase C and submitted their code, even if system detections were wrong. We also observed a higher early detection ratio in phase A led to decreased average active time in phase C [Figure \ref{fig:phaseC}] generalizing scenario 1. Scenario 2 played out for 7 of the 9 remaining students, who all submitted the program with incomplete objective(s) after spending 12-56 minutes in this phase. 
\begin{figure}[]
\centering
\includegraphics[width=0.3\textwidth]{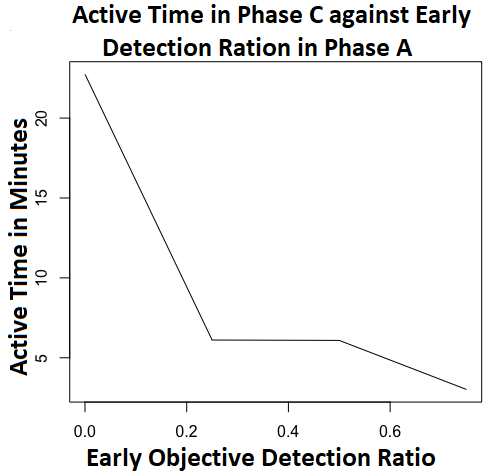}
\caption{Active Time in Phase C against Early Objective Detection Ratio in Phase A}
\end{figure}\label{fig:phaseC}

\subsubsection{Findings}
The results of our analysis showed that the active and idle time spent in Phase B and Phase C are associated with the quality of detection in Phase A. Correct objective detection in Phase A that led to a working solution pushed students to finish their attempts, making Phases B and C shorter. Whereas, incorrect objective detection in Phase A that led to a non-working solution decreased the idle time observed in Phase B and students worked more. However, the active time in such cases varied from student to student and depended on the extent to which objective detection went wrong. Our case studies presented in Section 6 demonstrated an indication of students’ significant reliance on the feedback. This reliance on the system interacted differently depending on whether the feedback was correct or incorrect, and whether or not the student code output appeared correct, and these differences are reflected in students’ responses or efforts in terms of active and idle time.

\section{Discussion}
The multi-criteria evaluation revealed important insights into the strengths (guidance for students in the absence of a human tutor despite being fallible and a source of motivation that improves persistence) and weaknesses (fallibility/limitations) of the DDAIF system which could be applicable to similar data-driven feedback systems. The evaluation also investigates students' behavior in relationship with feedback accuracy which showed how over-reliance or lack of self-assessment can lead to a negative impact (increased time, partial correct/incorrect solution) during problem-solving. These insights reinforced the necessity of system evaluations like ours to ensure proper usage and future development of data-driven feedback systems. Based on the results of our evaluation, we would provide the following recommendations regarding the use of data-driven feedback systems and future research directions in the corresponding field:
\begin{enumerate}
    \item Data-driven feedback systems [specifically in the domain of programming] should undergo a multi-criteria system evaluation before large-scale deployment that considers fallibility and its impact along with other evaluation metrics.
    \item Mitigation steps should be planned and implemented based on observed impacts of incorrect feedback for such systems. In the case of DDAIF, the observed impacts possibly resulted from students’ high reliance on the system and lack of self-evaluation. To mitigate such impact, self-evaluation should be promoted. Additionally, the system fallibility should be communicated effectively to students so that they do not lose trust in the system and can leverage the benefits without falling victim to incorrect feedback.
    \item Since, systems similar to DDAIF do not adapt to new programming behavior, to increase the correctness of such systems, an iterative process should be implemented for integrating new behaviors that may arise from diversity in student approaches. Semantics-based feedback generation should also be explored to reduce faulty feedback due to syntactic dissimilarity between a new program and historical correct programs.
\end{enumerate}
\section{Conclusion}
The contributions of this paper are 1) A multi-criteria evaluation mechanism for data-driven feedback systems applicable for all domains; 2) Easily adaptable and extendable data-driven methods to address each criterion of the evaluation; 4) Strengths and limitations of data-driven feedback systems; and 3) Important insights on relationships between the correctness of provided feedback and impact on students' response derived from the evaluation of DDAIF and recommendations applicable for similar data-driven feedback systems. In our future work, we plan to explore these impacts in larger controlled studies and on other programming tasks, and we also plan to explore how we can adapt our system to balance students' understanding of their own code with reliance on feedback, to promote learning.



\section*{Acknowledgment}
This material is based upon work supported by the National Science Foundation under Grant No. xxxxxxx.

\ifCLASSOPTIONcaptionsoff
  \newpage
\fi



%
\bibliographystyle{IEEEtran}
\bibliography{IEEEabrv,IEEEexample}

%

\begin{IEEEbiography}[{\includegraphics[width=1in]{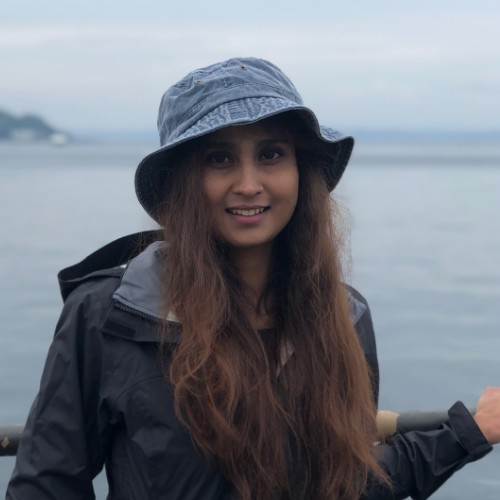}}]{Preya Shabrina}
Preya Shabrina is a Ph.D. Student in the Department of Computer Science at North Carolina State University (NCSU). Her research focuses on applying machine learning and data driven techniques to evaluate and improve automated supports integrated within Intelligent Tutoring Systems (ITS).
\end{IEEEbiography}
\vskip -3\baselineskip plus -1fil
\begin{IEEEbiography}[{\includegraphics[width=1in]{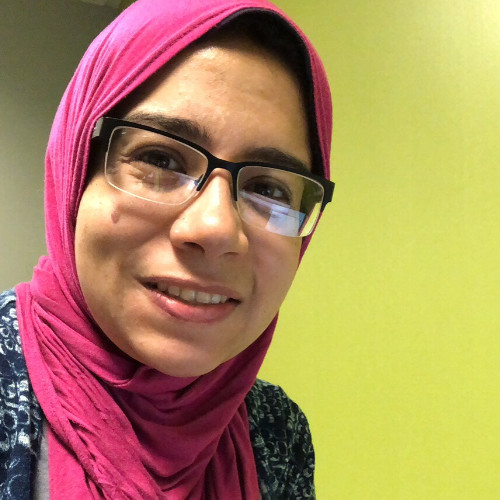}}]{Samiha Marwan}
Samiha Marwan is a research assistant in the HINTS Lab at NCSU, working in developing intelligent support features in block-based programming languages to improve novices' cognitive and affective outcomes.
\end{IEEEbiography}
\vskip -3\baselineskip plus -1fil
\begin{IEEEbiography}[{\includegraphics[width=1in]{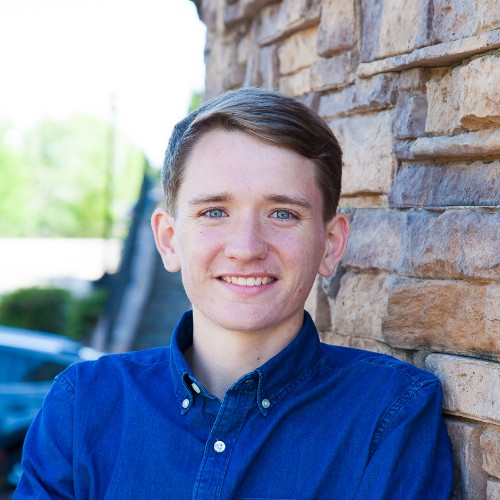}}]{Andrew Bennison}
Andrew Bennison is an undergraduate research assistant majoring in Computer Science at NCSU. He is personally interested in researching improvements in teaching for novice programmers learning how to code.
\end{IEEEbiography}
\vskip -3\baselineskip plus -1fil
\begin{IEEEbiography}[{\includegraphics[width=1in]{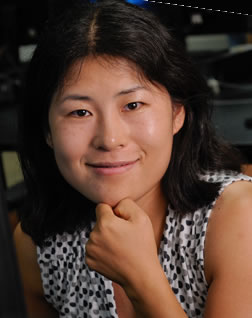}}]{Min Chi}
Dr. Min Chi is an Associate Professor in the Department of Computer Science at NC State University. Dr. Chi research primarily focuses on applying machine learning and data mining methods to improve human learning and exploring new machine learning and data mining challenges posed by learning and social science.
\end{IEEEbiography}
\vskip -3\baselineskip plus -1fil
\begin{IEEEbiography}[{\includegraphics[width=1in]{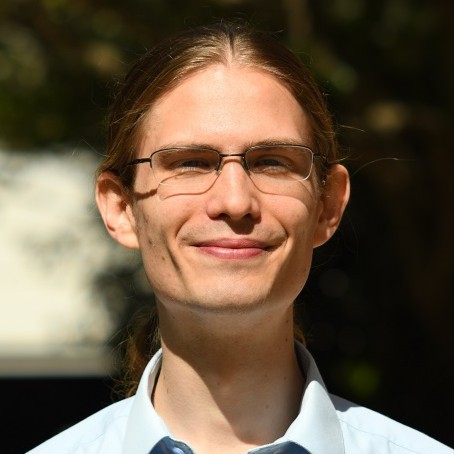}}]{Thomas Price}
Thomas Price is an Assistant Professor of Computer Science at NC State University. He directs the Help through INTelligent Support (HINTS) Lab, which develops learning environments that automatically support students through AI and data-driven help features. 
\end{IEEEbiography}
\vskip -3\baselineskip plus -1fil
\begin{IEEEbiography}[{\includegraphics[width=1in]{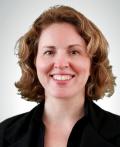}}]{Tiffany Barnes}
Tiffany Barnes is a Distinguished Professor of Computer Science at NC State University and co-Director for the STARS Computing Corps. Her research focuses on AI for education, educational data mining, serious games for education, computer science education, and broadening participation in computing education and research. She has served ACM SIGCSE (Symposium Chair 2018, Program Chair 2017, Board 2011-2016), IEEE Special Technical Community on Broadening Participation (Chair, and founder of the RESPECT conference (2015-present)), the International Educational Data Mining Society (EDM chair 2016, board 2011-present), STARS Computing Corps (Co-Director 2006-present, Celebration Chair 2011, 2015), Foundations of Digital Games (Program Chair 2014), the International Society for AI in Education (Board 2016-Present), and  IEEE Transactions on Learning Technologies (Assoc. Editor 2016-Present). 
\end{IEEEbiography}

\end{document}